\documentclass[prl,reprint,aps,amsmath,amssymb,twocolumn,superscriptaddress,longbibliography]{revtex4-1}
\usepackage{graphicx,bm,color,appendix}
\usepackage[colorlinks,bookmarks=true,citecolor=blue,linkcolor=blue,urlcolor=blue]{hyperref}
\usepackage{times}
\usepackage{amsmath,amssymb}
\usepackage{ytableau} 
\usepackage{comment}
\usepackage{appendix}

\usepackage{xcolor}
\definecolor{darkblue}{rgb}{0.,0.,0.4}
\definecolor{darkred}{rgb}{0.5,0.,0.}
\definecolor{BlueViolet}{RGB}{138,43,226}
\definecolor{SkyBlue}{RGB}{30,144,255}
\definecolor{DarkGreen}{RGB}{0,100,0}

\usepackage[normalem]{ulem}
\usepackage{comment}

\begin{document}

\title{ 
Reclaiming the Lost Conformality in a non-Hermitian Quantum 5-state Potts Model
}

\author{Yin Tang}
\affiliation{Department of Physics, School of Science, Westlake University, Hangzhou 310030, China }
\affiliation{School of Physics, Zhejiang University, Hangzhou 310058, China}

\author{Han Ma}
\affiliation{Perimeter Institute for Theoretical Physics, Waterloo, Ontario N2L 2Y5, Canada}

\author{Qicheng Tang}
\affiliation{Department of Physics, School of Science, Westlake University, Hangzhou 310030, China }

\author{Yin-Chen He}
\email{yhe@perimeterinstitute.ca}
\affiliation{Perimeter Institute for Theoretical Physics, Waterloo, Ontario N2L 2Y5, Canada}

\author{W. Zhu}
\email{zhuwei@westlake.edu.cn}
\affiliation{Department of Physics, School of Science, Westlake University, Hangzhou 310030, China }

\begin{abstract}

Conformal symmetry, emerging at critical points, can be lost when renormalization group fixed points collide. Recently, it was proposed that after collisions, real fixed points transition into the complex plane, becoming complex fixed points described by complex conformal field theories (CFTs). Although this idea is compelling, directly demonstrating such complex conformal fixed points in microscopic models remains highly demanding. Furthermore, these concrete models are instrumental in unraveling the mysteries of complex CFTs and illuminating a variety of intriguing physical problems, including weakly first-order transitions in statistical mechanics and the conformal window of gauge theories. In this work, we have successfully addressed this complex challenge for the (1+1)-dimensional quantum $5$-state Potts model, whose phase transition has long been known to be weakly first-order. By adding an additional non-Hermitian interaction, we successfully identify two conjugate critical points located in the complex parameter space, where the lost conformality is restored in a complex manner. Specifically, we unambiguously demonstrate the radial quantization of the complex CFTs and compute the operator spectrum, as well as new operator product expansion coefficients that were previously unknown.

\end{abstract}

\maketitle

\textit{Introduction.---}Conformal field theory (CFT) provides a comprehensive framework for understanding continuous phase transitions and associated critical phenomena~\cite{Cardy_book,yellowbook}. However, there are scenarios where conformal symmetry is lost, particularly when conformal fixed points collide and disappear in real physical space~\cite{Kaplan2009}, leading to first-order transitions. A recent intriguing theory posits that after such collisions, these conformal fixed points relocate to the complex plane~\cite{Wang2017,Gorbenko2018a,Gorbenko2018b}, where conformal symmetry is reestablished in a complex manner. These phenomena give rise to what are now known as complex CFTs~\cite{Gorbenko2018a,Gorbenko2018b}. Complex CFTs, distinguished by unique properties like complex scaling dimensions, represent a fundamentally new category of non-unitary CFTs compared to well-known examples such as the 2D Lee-Yang singularity in minimal models \cite{leeyang1952,Cardy1985c,yellowbook}.

Beyond their theoretical significance, complex CFTs are pivotal in addressing many unresolved problems. They play an essential role in clarifying weakly first-order phase transitions, including the extensively studied deconfined phase transitions~\cite{Senthil2004a,Senthil2004b,Nahum2015,Wang2017,Gorbenko2018a,Serna2019,Ma2020,Nahum2022,Zhou2023}, and are key in accurately determining the conformal window of critical gauge theories~\cite{Kaplan2009,Gorbenko2018a,Benini2020,Antipin2020,Uetrecht2023}. However, most efforts to address these problems have been concentrated on weakly first-order transitions in the unitary parameter space, where at best, one can observe an approximate conformal symmetry as well as walking (also called  pseudo-critical) renormalization group (RG) flow influenced by the nearby complex fixed points \cite{Ma2019,Roberts2021,Zhou2023}. There is a pressing need to harness the full potential of complex conformality, which involves studying the complex fixed points in microscopic models extended to the non-unitary parameter space. Additionally, such microscopic realizations are invaluable for understanding complex CFTs themselves, whose properties remain enigmatic, even in the simplest examples, due to the limitation of existing theoretical tools~\cite{Faedo2020,Benedetti2021,Giombi2020,Benini2020,Han2023,Uetrecht2023,Gorbenko2018a,Gorbenko2018b,Haldar2023,Wiese2023}.

\begin{figure}
\includegraphics[width=0.45\textwidth]{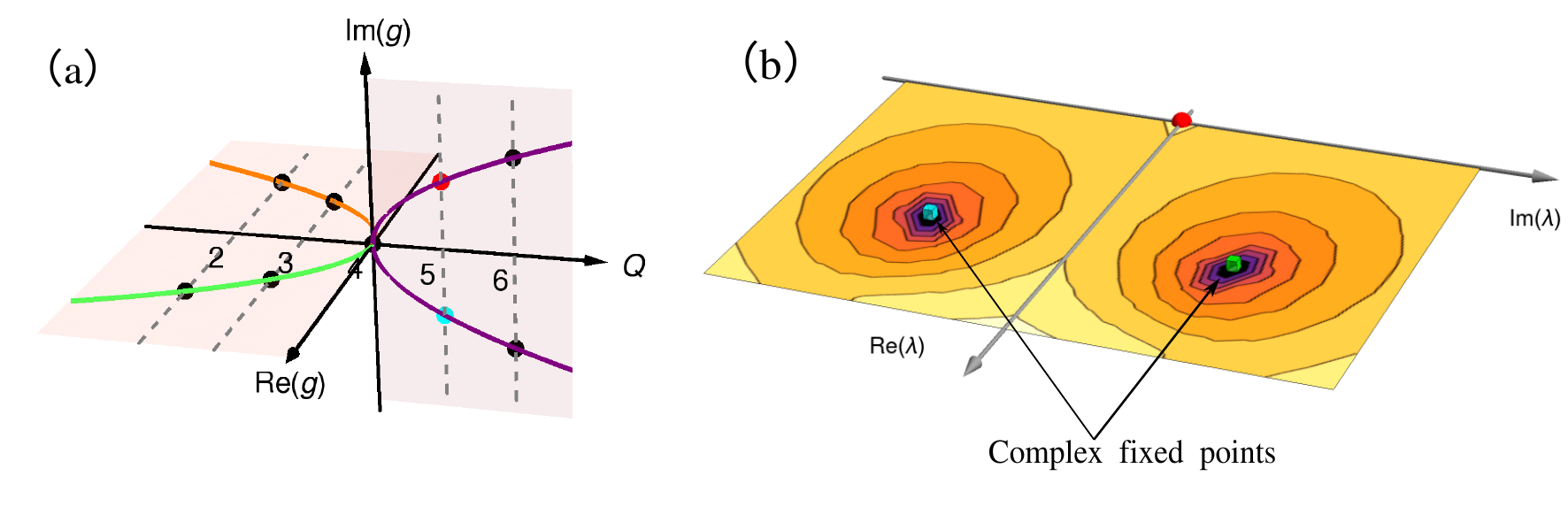}
\caption{\label{fig:drawing} 
(a) The complex CFT scenario for the Q-state Potts model: The phase transitions are second-order for $Q\le 4$, while weakly first-order for $Q>4$ because the location of the critical branches move into the complex parameter plane. (b) Illustration of phase transitions of the non-Hermitian 5-state Potts model $H_{\text{NH-Potts}}(J,h,\lambda)$ on the self-dual plane ($J/h=1$): Two complex fixed points located at $\lambda_c\neq0$ (green and cyan squares), which are determined via conformal perturbation calculation (see main text).
The transition point of original 5-state Potts model is marked by the red dot. 
}
\end{figure}

In this paper, we successfully tackle the complex challenge for the two-dimensional 5-state Potts model, a model long known for its subtly nuanced weakly first-order transition~\cite{Baxter1973,Nienhuis1979,Cardy1980,Nauenberg1980,Nienhuis1980,Nienhuis1981,Wu1982,Buddenoir1993}.  Specifically, we introduce a (1+1)-dimensional non-Hermitian quantum 5-state Potts model and identify its two conjugate complex fixed points.  At the complex critical point, employing the state-operator correspondence~\cite{Cardy1984,Blote1986,Cardy1985a}, we observe clear evidence of emergent conformal (i.e., Virasoro) invariance and extract the complex scaling dimensions of 11 low-lying Virasoro primary fields \cite{CARDY1986,Affleck1988}. Additionally, we calculate 9 distinct operator product expansion (OPE) coefficients involving primary fields, most previously unknown. Our results not only directly confirm the complex origin of the weakly first-order phase transition in the two-dimensional 5-state Potts model, but also represent a solid advancement towards unraveling three-dimensional mysteries, such as deconfined phase transitions and critical gauge theories, using the recently proposed fuzzy sphere regularization~\cite{ZHHHH2022,Zhou2023}.

\begin{figure}
\includegraphics[width=0.48\textwidth]{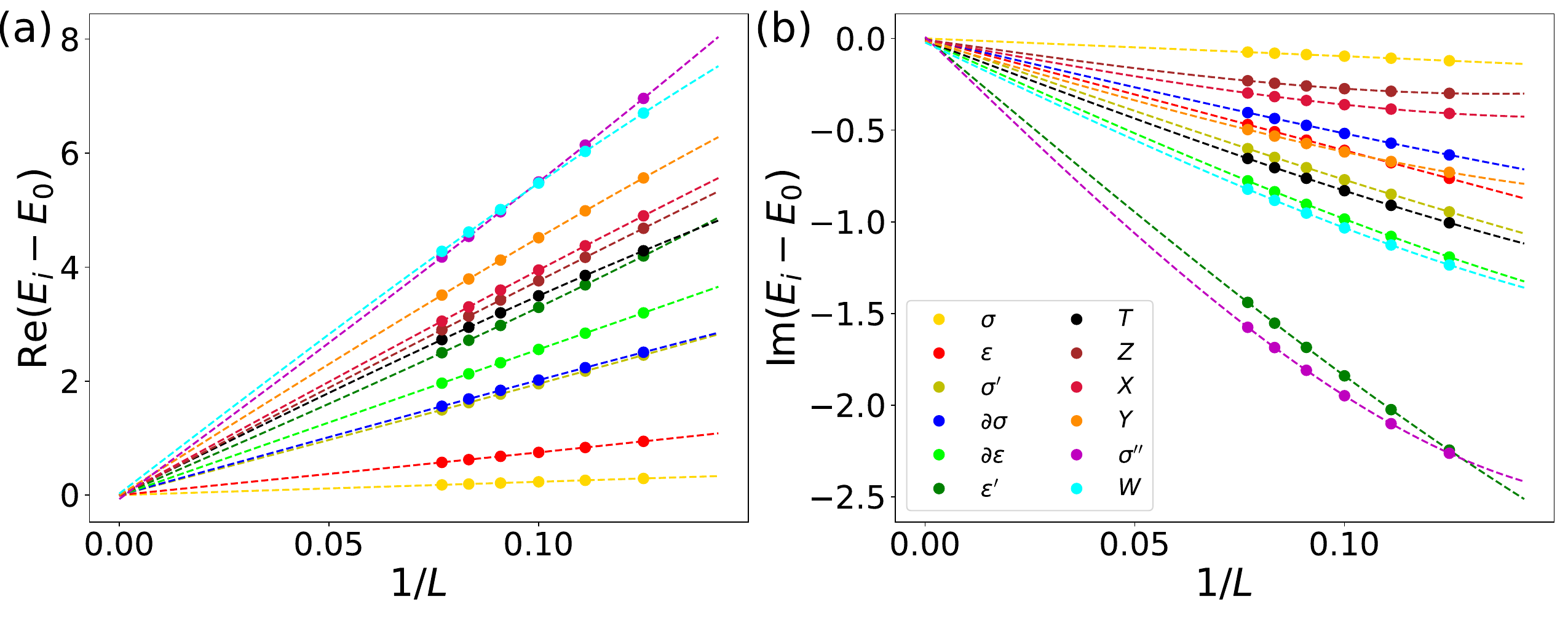}
\caption{\label{fig:scale_gap} 
Finite-size scaling of the low-lying excitation energy gaps (distinguished by colors) at the complex critical point, where (a) real and (b) imaginary part of energy gaps identically scales to zero. 
}
\end{figure}

\textit{Model.---}The Hamiltonian of the $Q$-state quantum Potts model is \cite{Wu1982} 
\begin{equation} \label{eq:h_potts}
H_0(J,h) = - \sum_{i=1}^{L} \sum_{k=1}^{Q-1} [J ( \sigma_i^{\dagger}  \sigma_{i+1})^k +h  \tau_i^k ],
\end{equation}
where the spin shift operator $\hat{\tau}$ and phase operator $\hat{\sigma}$ respectively changes the local spin degrees of freedom as $\hat{\tau} |n\rangle = | (n+1) \text{ mod } Q\rangle$ and $\hat{\sigma} |n\rangle = e^{2\pi n i/Q} |n\rangle$. The Hamiltonian is invariant under the spin permutation $S_Q$ symmetry, has two phases, i) an ordered phase at $J> h $ that spontaneously breaks $S_Q$ symmetry, and ii) an disordered phase at $J< h$ that respects $S_Q$ symmetry. The order-disorder transition occurs at $J=h$, with its precise position determined by the Kramers-Wannier duality. It is well established that the phase transition is continuous for $Q\le 4$ but is first-order for $Q>4$~\cite{Baxter1973,Nienhuis1979,Cardy1980,Nauenberg1980,Nienhuis1980,Nienhuis1981,Buddenoir1993}. Notably, for $Q$ just above 4, such as $Q=5$, the first-order transition is very weak characterized by a large correlation length and a small energy gap.

To gain a deeper insight into the phase transition in the Potts model, one can examine a more comprehensive parameter space by introducing an extra $S_Q$ and Kramers-Wannier duality invariant term into the Hamiltonian. In this work, we propose to consider the following term:
\begin{equation}
\begin{aligned}
\label{eq:h_int}
H_1 (\lambda) =\lambda \sum_{i=1}^{L} \sum_{k_1,k_2=1}^{Q-1}  &  [(\tau_i^{k_1} +\tau_{i+1}^{k_1}) (\sigma_i^{\dagger} \sigma_{i+1})^{k_2}
\\
& + (\sigma_i^{\dagger} \sigma_{i+1})^{k_1} (\tau_i^{k_2} +\tau_{i+1}^{k_2}) ].
\end{aligned}
\end{equation}
As illustrated in Fig.~\ref{fig:drawing}(a), by tuning $\lambda$ there are indeed two (real) fixed points for $Q<4$,  one is attractive corresponding to the $Q$-state Potts CFT, and the other is repulsive corresponding to the tri-critical $Q$-state Potts CFT~\cite{Dotsenko1984a,Dotsenko1984b,Zamolodchikov1985,Zamolodchikov1987}. At $Q=4$, these two fixed points merge into a single fixed point, corresponding to an orbifold free boson CFT~\cite{Dijkgraaf1989,Cardy1980}. For $Q>4$, the two fixed points collide and disappear in the real axis of $\lambda$,  the phase transition becomes first-order rather than continuous \cite{Baxter1973,Nienhuis1979,Cardy1980,Nauenberg1980,Nienhuis1980,Nienhuis1981,Wu1982,Buddenoir1993,Iino2019,D2023}. Nevertheless, by performing a naive analytical continuation of the $\beta$-function to the complex coupling $\lambda$, the previously disappeared fixed point will reemerge. It is proposed that these complex fixed points are still conformal, and are responsible for the weakly first-order transition for $Q$ slightly larger $4$~\cite{Wang2017,Gorbenko2018b}.

In the following, we will directly confirm the complex fixed point proposal by considering the non-Hermitian Potts model, $H_{\text{NH-Potts}} (J,h,\lambda) = H_0(J,h) + H_1 (\lambda)$. 
Specifically, we will identify the complex fixed points for $Q=5$, and study the corresponding complex CFT through the state-operator correspondence. We also note that similar lattice construction has been studied in Potts model to realize tri-critical Ising/3-state Potts (real) fixed points \cite{O2018,O2020b}.

\begin{figure*}
\includegraphics[width=0.75\textwidth]{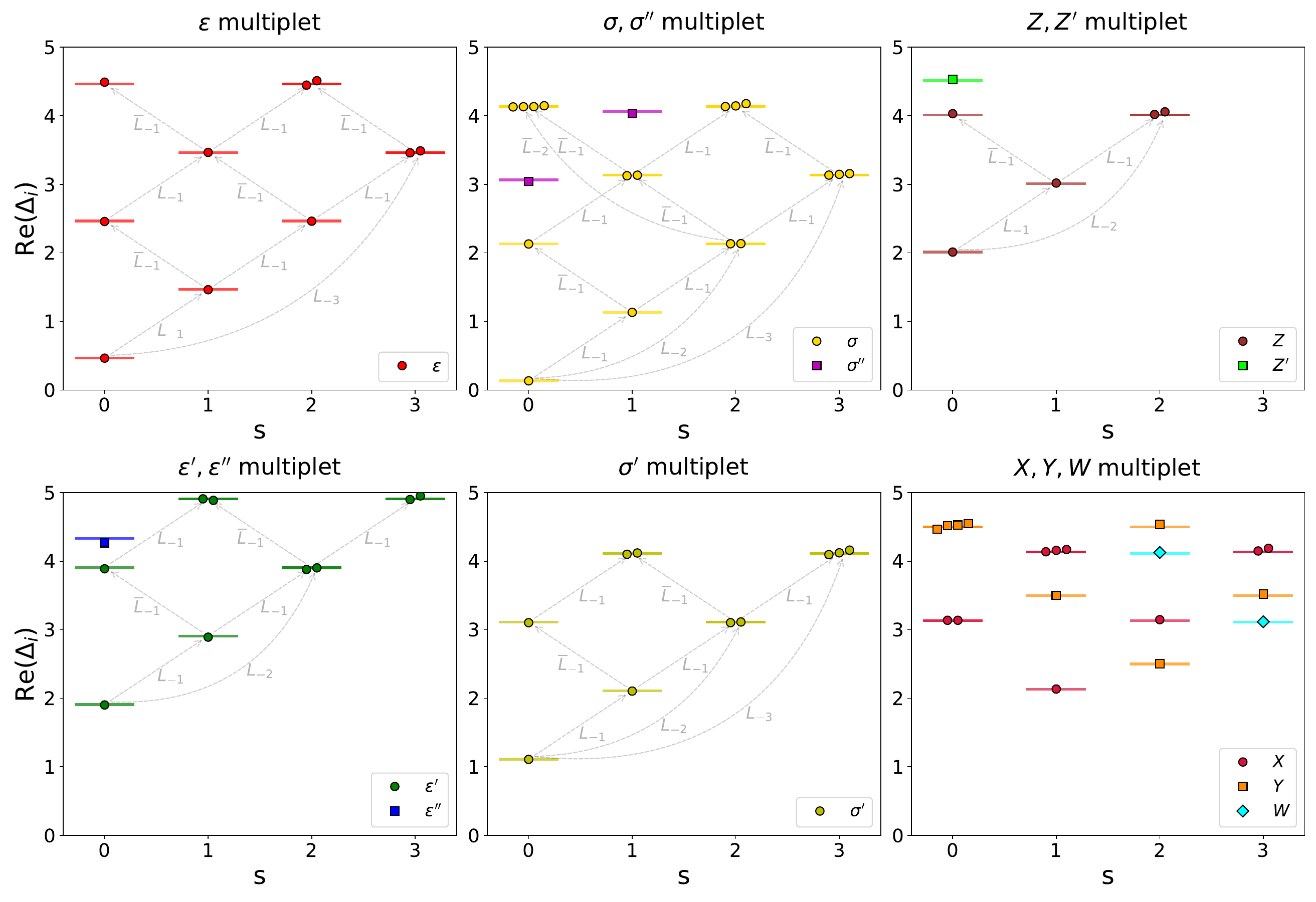}
\caption{\label{fig:con_fam} Conformal multiplet for 11 low-lying Virasoro primary operators: real part of scaling dimension Re$(\Delta)$ versus Lorentz spin $s$. Different symbols and colors label different conformal towers. The spectrum is calibrated by setting the scaling dimension of energy momentum tensor $\Delta_T=2$. The dots are results from the 5-state Potts model and short lines are prediction from the analytical continuation of the Coulomb Gas partition function \cite{Gorbenko2018b}. The translucent arrows denote Virasoro generators connecting different states.
}
\end{figure*}

\begin{table}
\caption{\label{tab:ope_dim_primary} Operator scaling dimensions for 11 low-lying Virasoro primary fields identified through the state-operator correspondence. $s$ represents the Lorentz spin quantum number. `$S_Q$ Rep' presents the young diagrams for irreducible representations of permutation group. The numerical extrapolated data from the non-Hermitian 5-state Potts model is compared with the prediction based on analytical continuation \cite{Gorbenko2018b}. 
}

\ytableausetup{boxsize=0.5em}

\begin{ruledtabular}
\begin{tabular}{ccccc}
Operator  & s & $S_Q$ Rep& complex CFT & non-Hermitian 5-Potts  \\ \hline

 $\epsilon$ &  $0$   & $\ydiagram{5}$ & $0.466-0.225i$ &  $0.463-0.224i$   \\

$\epsilon'$ & $0$ & $\ydiagram{5}$ & $1.908-0.599i$ &  $1.900-0.598i$ \\

$\epsilon''$  & $0$ & $\ydiagram{5}$ & $4.328-1.123i$ &  $4.324-1.122i$ \\ \hline

 $\sigma$  & $0$ & $\ydiagram{4,1}$ & $0.134-0.021i$ &  $0.133-0.021i$  \\

$\sigma'$  & $0$ & $\ydiagram{4,1}$ & $1.111-0.170i$ &  $1.108-0.170i$ \\

$\sigma''$  & $0$ & $\ydiagram{4,1}$ & $3.065-0.470i$ &  $3.065-0.470i$  \\ \hline

 $Z$  & $0$ & $\ydiagram{3,2}$ & $2.012+0.305i$ &  $2.016+0.304i$ \\

$Z'$  & $0$ & $\ydiagram{4,1}$ & $4.512+0.688i$ &  $4.536+0.698i$ \\

$X$  & $1$ & $\ydiagram{3,1,1}$ & $2.134+0.286i$ &  $2.138+0.286i$ \\

$Y$  & $2$ & $\ydiagram{3,2}$ & $2.500+0.230i$ &  $2.503+0.230i$\\

$W$  & $3$ & $\ydiagram{3,1,1}$ & $3.111+0.136i$ &  $3.110+0.137i$ \\

\end{tabular}
\end{ruledtabular}
\end{table}

\textit{Critical points in the complex parameter space.---}
We begin by analyzing the phase transition point of $H_{\text{NH-Potts}}(J,h,\lambda)$. 
Firstly, we assume the microscopic model $H_{\text{NH-Potts}}(J,h,\lambda)$ realizes an effective model $H_{\text{CFT}}(g_{\epsilon'})=H_{\text{CFP}} + g_{\epsilon'} \int \mathrm{d}x \cdot \epsilon'(x) + ...$, where $\epsilon'$ is the most relevant $S_Q$ singlet preserving duality and $g_{\epsilon'}$ is the coupling constant. Tuning parameters to hit the critical value $(J_c,h_c,\lambda_c)$, $H_{\text{NH-Potts}}(J_c,h_c,\lambda_c)$ realizes a complex CFT fixed point $H_{\text{CFP}}$ with vanishing small $g_{\epsilon'}$. Accordingly, one can compare the energy spectrum of $H_{\text{NH-Potts}}(J,h,\lambda)$ with those of $H_{\text{CFT}}(g_{\epsilon'})$, and the minimum of $g_{\epsilon'}$ should point to the location of critical point of  $H_{\text{NH-Potts}}$ \cite{Lao2023}, as shown in Fig. \ref{fig:drawing}(b). Following this process, we obtain $(J_c=h_c=1,\lambda_c = 0.079+0.060i)$ and its complex conjugated point $ (J_c=h_c=1,\bar{\lambda}_c =0.079-0.060i)$, which we determine as the critical points (Supple. Mat. Sec. B \cite{sm}). 
Given that the Hermitian breaking term $\lambda_c$ is very small, the weakly first-order transition point in the original 5-state Potts model (at $J=h=1, \lambda=0$) is notably close to the complex critical points.

Next, at the critical point, according to the expectation of CFT the energy spectrum should follow the scaling form  
\begin{align}
 (E_n - E_0) & = \frac{2\pi v}{L} \Delta_n + ... 
\end{align}
where $L$ is the length of spin chain, $v$ is a non-universal normalization factor, and $E_0$ is the ground state energy \cite{Cardy1984,Blote1986,Cardy1985a}. The `$...$' stands for finite-size corrections from irrelevant operators (see Supple. Mat. Sec. C \cite{sm}). It is worth noting that the eigenenergies $E_n$ and scaling dimensions $\Delta_n$  of the CFT operator are complex values for complex CFT, in sharp contrast to unitary CFT and non-unitary CFT such as Lee-Yang. As shown in Fig.~\ref{fig:scale_gap} (a-b), by sitting at the phase transition point the excited energy gaps simultaneously scales to vanishing small values for both real and imaginary part, signaling the criticality. 
Additionally, the ground state energy density $E_0/L$ can give rise to the central charge $c \approx 1.1405-0.0224i$ (details see Supple. Mat. Sec. D \cite{sm}), close to the exact result from analytical continuation~\cite{Gorbenko2018b},
\begin{equation}  
\begin{aligned}  
c_{\textrm{5-Potts}} & = 7-\frac{12\pi}{2\pi + i \log(\frac{3+\sqrt{5}}{2})} - \frac{ 3i\log(\frac{3+\sqrt{5}}{2})}{\pi} \nonumber \\ &\approx 1.1375 - 0.0211i.
\end{aligned}  
\end{equation}

\textit{Operator spectrum.---}
Generally, a key feature of the CFT is the state-operator correspondence~\cite{Cardy1984, CARDY1986}, i.e. the eigenstate $|\phi\rangle$ of the CFT quantum Hamiltonians on a cylinder $\mathbb{S}^{d-1} \times \mathbb{R}^1$ has one-to-one correspondence with the CFT operator $\hat \phi$,
which allows to access the conformal data such as the scaling dimensions of CFT operators. 
Previously, the state-operator correspondence has been explicitly shown in CFTs with real scaling dimensions, such as the 2D Ising \cite{CARDY1986,Milsted2017,Zou2018,Zou2020b} and non-unitary Lee-Yang CFT \cite{Gehlen1991,Itzykson1986}.
Next, we will explicitly demonstrate this correspondence holds in the current model $H_{\text{NH-Potts}}(J_c,h_c,\lambda_c)$, even though it is non-Hermitian.

Fig.\ref{fig:con_fam} depicts the obtained operator spectra at the critical point, which are grouped into different conformal families in subfigures. The results have been extrapolated to the thermodynamic limit ($L \to \infty$) and we have rescaled the full spectra by setting energy-momentum tensor to be $\Delta_T=2$.
Here we plot the real-part of scaling dimensions (the imaginary part see \cite{sm}). 
The spectra shows unique and distinguishable features.
First of all, in each conformal family, the eigenstate with lowest real part is identified as Virasoro primary field, and higher energy states (associated with descendant fields) are nearly integer spacing separating from the primary field. For example, for a primary operator $\hat \phi$ with scaling dimension $\Delta_{\phi}$ and quantum number $(s,s_Q)$ (here $s$ is the  Lorentz spin, and $s_Q$ is spin permutation symmetry quantum number), its descendants can be represented as $L_{-\mu_1} \cdots L_{-\mu_m} \overline{L}_{-\nu_1} \cdots \overline{L}_{-\nu_n} \hat \phi$ ($1\leq \mu_1 \leq \cdots \leq \mu_m$, $1\leq \nu_1 \leq \cdots \leq \nu_n$), with scaling dimension ($\Delta_{\phi}+\sum^{m}_{i=1} \mu_i +\sum^{n}_{j=1} \nu_j$) and quantum number $(s+\sum^{m}_{i=1} \mu_i -\sum^{n}_{j=1} \nu_j,s_Q)$ \cite{sm}. 
Importantly, the degeneracy of the descendants satisfy with the expectation of Verma module. Note that applying Virasoro generators to some specific states might lead to null states, e.g. $L_{-2} |\epsilon\rangle$ and $L_{-3} |\epsilon'\rangle$ \cite{sm}. 
All of the above features demonstrate the emergent conformal symmetry regarding the identified critical point of $H_{\text{NH-Potts}}(J_c,h_c,\lambda_c)$.
As far as we know, this hidden complex conformality has not been demonstrated by any quantum Hamiltonian before.
Moreover, comparing with the unitary 5-state Potts model (see Fig. S3 \cite{sm}), the spectra at the weakly first-order transition point deviates from the CFT tower structure, implying that the transition point there is not exactly continuous.

Having clarifying the emergent conformal symmetry at the complex fixed point, we further investigate the scaling dimensions of the identified
primary operators, as listed in Tab.\ref{tab:ope_dim_primary}. Crucially, we find  $\epsilon'$ ($\Delta_{\epsilon'}\approx 1.908-0.599i$), which controls the RG flow around the complex fixed points \cite{Gorbenko2018b}. 
Moreover, in comparison with the CFT data from analytical continuation \cite{Gorbenko2018b,Di1987}, the quantitative agreement has been confirmed, i.e. for all 11 Virasoro primary operators that we have identified, the discrepancy is less than $1\%$ for the real and $2\%$ for the imaginary part. 
Even more remarkably, we have checked all low-lying states with Re$\Delta_n< 5, s \le 3$, and they perfectly match the CFT spectrum (for both primary and descendant operators), without any extra or missing operator (see Tab. S2-S11 \cite{sm}).

\textit{Correlation functions and OPE coefficients.---}
Next we turn to access the OPE coefficients. 
In general, any local lattice operator $ O$ can be expanded by the linear combination of CFT operators: $O=\sum_\alpha c_\alpha \hat \phi_\alpha$, where the summation is over infinite primary and descendant operators and $c_\alpha$ are some non-universal coefficients. 
For example, spin operator $\sigma_i$ should involve operator content of CFT operators with the same $S_Q$ quantum number:
\begin{equation}
\begin{aligned}
O_{\sigma} =  \sigma_i =(a_{\sigma} \cdot \hat{\sigma} + \text{desc.}) + (b_{\sigma'} \cdot \hat{\sigma}'  + \text{desc.}) +\cdots,
\end{aligned}
\end{equation}
where `$\cdots$' represents other CFT operators with higher scaling dimensions. 
This operator content decomposition can be further inspected by corresponding two-point correlators \cite{sm}.

Then we calculate correlation functions to extract corresponding OPE coefficients with the help of  state-operator correspondence~\cite{Zou2020a}.
For example, the OPE $C_{\alpha \sigma \beta}$ can be computed from (Supple. Mat. Sec. E \cite{sm})
\begin{equation}
\begin{aligned}
\label{eq:ope}
&\sqrt{ \frac{  {_L}\langle \alpha| O_{\sigma}|\beta\rangle_{R} {}_{L}\langle \beta| O_{\sigma}|\alpha\rangle_{R}  }{ _{L}\langle  0 | O_{\sigma}|\sigma\rangle_{R} {}_{L}\langle \sigma| O_{\sigma}|0\rangle_{R} }  } 
 \approx  C_{\alpha \sigma \beta} + \frac{c_1}{L^{\Delta_{\sigma'}-\Delta_{\sigma}}} \\& \,\,\,\,\,\,\,\,\,\,\,\,\,\,\,\,\,\,\,\,
 + \frac{c_2}{L^{2(\Delta_{\sigma'}-\Delta_{\sigma})}}+ \frac{c_3}{L^2}+O(\frac{1}{L^{\Delta_{\sigma'}-\Delta_{\sigma}+2}}),
\end{aligned}
\end{equation}
where subscripts $L$ and $R$ represents left and right biorthogonal eigenstates of $H_{\text{NH-Potts}}(J_c,h_c,\lambda_c)$ and $c_{1,2,3}$ are non-universal coefficients. The proposed form Eq. \ref{eq:ope} is to remove the gauge redundancy of left and right eigenstates. 
Subsequently a finite-size extrapolation is performed to  eliminate the contribution from higher primary and descendant fields. 
Similarly, other OPE coefficients involving $\epsilon (\epsilon')$  can be obtained by using duality-odd (even) operator $O_\epsilon ( O_{\epsilon'})$ \cite{sm}.
Tab.\ref{tab:ope} summarizes the obtained 9 different OPE coefficients. One typical feature distinguished from unitary CFTs is the OPE coefficients are complex values. Crucially, before the current work, only 4 of these OPE coefficients could be obtained through an analytical continuation from previous results of minimal models \cite{Poghossian2013}.


\begin{table}
\caption{\label{tab:ope} 
OPE coefficients from the non-Hermitian 5-state Potts model,  in comparison with analytical continuation from minimal models \cite{Poghossian2013,Gorbenko2018b}. 
}
\begin{ruledtabular}
\begin{tabular}{c|cc}
$\text{OPE} $ &  \text{non-Hermitian 5-Potts} & \text{Analytical Continuation} \\ \hline

 $C_{\epsilon \epsilon \epsilon'}$ & $0.8781(72)-0.1433(21)i$ & $0.8791-0.1404i$ \\

 $C_{\epsilon' \epsilon' \epsilon'}$ & $2.2591(77)-1.1916(39)i$ & $2.2687-1.1967i$ \\

 $C_{\epsilon \epsilon' \epsilon''}$ & $0.804(3)-0.220(11)i$  & $0.8318-0.2027i$ \\

  $C_{\epsilon'' \epsilon'' \epsilon'}$ & $3.886(96)-3.369(33)i$ & $3.9261-3.3261i$ \\

$C_{\sigma \sigma \epsilon'}$ & $0.0658(15)+0.0513(10)i$  & NA \\

 $C_{\sigma \sigma \epsilon}$ & $0.7170(6)+0.1558(1)i$ & NA  \\

$C_{\sigma \sigma' \epsilon}$&$  0.7520(15)-0.0831(6)i$ & NA \\

$C_{\sigma \sigma' \epsilon'}$&$ 0.6062(4)+0.1664(10)i$ & NA \\

 $C_{\sigma \sigma'' \epsilon'}$&$ 0.6709(45)-0.1250(21)i$ & NA \\

\end{tabular}
\end{ruledtabular}
\end{table}

\textit{Summary and discussion.---}We have constructed a non-Hermitian 5-state quantum Potts model, which realizes complex fixed points described by an $S_5$ symmetric complex CFT. At the critical point, we provide compelling evidence for emergent conformal symmetry and present a clear demonstration of state-operator correspondence in complex CFT. Specifically, we have identified the conformal data, including the scaling dimensions of 11 Virasoro primary operators and 9 associated OPE coefficients with high accuracy. These findings are significant in several respects. First, our microscopic calculations shed light on the existence of complex CFT and its corresponding complex criticality. Second, they directly verify the proposal that the first-order regime $(Q>4)$ of the two-dimensional Potts model is in proximity to complex fixed points, paving the way for future investigations into similar models, including gauge theories below the conformal window and deconfined quantum critical points. Third, most of the conformal data computed in this work, such as OPE coefficients, have not been obtained non-perturbatively before. This essential information should facilitate other methods for solving complex conformal data \cite{bootstrap_rmp,Jacobsen2018,Nivesvivat2021,jacobsen2023,nivesvivat2023}. Finally, this work should also inspire further exploration into quantum critical phenomena in non-Hermitian Hamiltonians.

\textit{Note added.---} During the preparation of this work, we become aware of an independent study (arXiv.2402.10732 \cite{jacobsen2024lattice}) on the complex fixed points of classical 5 state Potts model.

\textit{Acknowledgment.---}	
We acknowledge stimulating discussion with Slava Rychkov.
This work was supported by NSFC No. 92165102 and the National Key R$\&$D program No. 2022YFA1402204. 
Research at Perimeter Institute is supported in part by the Government of Canada through the Department of Innovation, Science and Industry Canada and by the Province of Ontario through the Ministry of Colleges and Universities. 

\bibliography{ref}

\clearpage

\appendix

\begin{widetext}

\begin{center}
\textbf{Supplementary materials for ``Reclaiming the Lost Conformality in a non-Hermitian Quantum 5-state Potts Model ''}    
\end{center}

\setcounter{subsection}{0}
\setcounter{equation}{0}
\setcounter{figure}{0}
\renewcommand{\theequation}{S\arabic{equation}}
\renewcommand{\thefigure}{S\arabic{figure}}
\renewcommand{\thetable}{S\arabic{table}}
\setcounter{table}{0}

\tableofcontents

\vspace{2\baselineskip} 

This supplementary material contains the following materials to support the discussion in the main text. 
In Sec. A, we firstly review two-dimensional complex conformal field theory (CFT). Then we discuss the process to determine the phase transition point using conformal perturbation theory in Sec. B. It also enables us to evaluate the finite-size correction to the energy spectrum, through which we could extract the complex scaling dimensions (Sec. C). 
In Sec. D, we try to extract the central charge at the critical point. In Sec. E, we present the extrapolated operator spectrum at the weakly first-order phase transition point of the original 5-state Potts model, for a comparison with the data shown in the main text. Finally, we identify the operator content of several local lattice operators associated with our model in Sec. F and employ these decomposition to evaluate some OPE coefficients in Sec. G.

\section{A. Review of 2D (complex) CFT}

Generally speaking, complex CFT describes the fixed points of complex RG flows  \cite{Gorbenko2018a}. In this case, the conformal data is no longer real. Importantly, the general properties of CFT are still satisfied, such as conformal symmetry, operator product expansion and crossing equations.
In this section, we review some basic knowledge of two dimensional complex CFT, in parallel to unitary CFTs.

The study of CFT focus on quantum field theory invariant under conformal transformation $x\rightarrow x'$, which transforms metric as \cite{yellowbook}

\begin{equation}
g_{\rho\sigma}'(x') \frac{\partial x'^{\rho}}{\partial x^{\mu}} \frac{\partial x'^{\sigma}}{\partial x^{\nu}} = \Lambda(x) g_{\mu\nu}(x)
\end{equation}

In general spacetime dimensions, conformal transformations usually include translations, dilations, rotations and special conformal transformations, with corresponding generators $P_{\mu}=-i\partial_{\mu}$, $D=-ix^{\mu}\partial_{\mu}$, $L_{\mu\nu} = i(x_{\mu}\partial_{\nu}-x_{\nu}\partial_{\mu})$ and $K_{\mu} =-i(2x_{\mu}x^{\nu}\partial_{\nu}-(x\cdot x)\partial_{\mu})$. It is more convenient to redefine these generators through
\begin{equation}
\begin{aligned}
J_{\mu\nu} &= L_{\mu\nu}
\\
J_{-1,0} & = D
\\
J_{-1,\mu} &= \frac{1}{2} (P_{\mu}-K_{\mu})
\\
J_{0,\mu} &=\frac{1}{2}(P_{\mu}+K_{\mu})
\end{aligned}
\end{equation}
Then one could verify that these novel generators $J_{\mu\nu}$ satisfy the following commutation relations:
\begin{equation}
[J_{\mu\nu},J_{\rho\sigma}] =i(\eta_{\mu\sigma}J_{\nu\rho}+\eta_{\nu\rho}J_{\mu\sigma}-\eta_{\mu\rho}J_{\nu\sigma}-\eta_{\nu\sigma}J_{\mu\rho})
\end{equation}
If we work on $d$-dimensional Euclidean space $\mathbb{R}^d$, the metric here becomes $\eta_{\mu\nu}=\text{diag}(-1,1,\cdots,1)$. In other words, the conformal group in $d$-dimensional Euclidean space is $SO(1,d+1)$.

In two dimensional Euclidean spacetime, if we consider an infinitesimal coordinate transformation $x'^{\mu} = x^{\mu}+\epsilon^{\mu}(x)$, the requirement of conformal transformation is reduced to Cauchy-Riemann condition \cite{yellowbook}:
\begin{equation}
\partial_1 \epsilon_1 = \partial_2 \epsilon_2, \quad \partial_1 \epsilon_2 = -\partial_2 \epsilon_1
\end{equation}
Thus, if we introduce complex coordinates via $z,\bar{z}=x^1 \pm ix^2$, each two dimensional conformal transformation corresponds to a holomorphic function $\epsilon(z)$ in some open set. In complex analysis, there are infinite number of analytical functions in the complex plane, leading to infinite conformal symmetry in two dimension CFTs.

Conformal invariance imposes strong constraint on physical properties. In 2D CFT, there includes a set of scaling operators called primary fields $\phi(z,\bar{z})$. Under an arbitrary conformal transformation $z\to \omega(z),\bar{z}\to\bar{\omega}(\bar{z})$, these field operators transform accordingly through
\begin{equation}
\phi'(\omega,\bar{\omega}) = \big{(}\frac{\mathrm{d}\omega}{\mathrm{d}z}\big{)}^{-h} \big{(}\frac{\mathrm{d}\bar{\omega}}{\mathrm{d}\bar{z}}\big{)}^{-\bar{h}} \phi(z,\bar{z})
\end{equation}
where $h,\bar{h}$ are their conformal weights. Then the scaling dimension and conformal spin of this operator are given by $\Delta=h+\bar{h}$ and $s=h-\bar{h}$ respectively. In the case of unitary CFTs, $\Delta$ takes real numbers. One significant difference is, scaling dimensions $\Delta$ take complex values in complex CFTs. Additionally, the two point correlation functions are completely determined by conformal symmetry to take the following form
\begin{equation}
\begin{aligned}
_{_L}\langle \phi_1(z_1,\bar{z}_1) \phi_2 (z_2,\bar{z}_2)\rangle_{_R} = \frac{\delta_{12}C_{12}}{(z_1-z_2)^{h_1+h_2}(\bar{z}_1-\bar{z}_2)^{\bar{h}_1+\bar{h}_2}}
\end{aligned}
\end{equation}
where the notion of left and right eigenvector $| \rangle_{L(R)}$ should be distinguished in the complex CFT which will be explained below. 

Another useful property of CFT is operator product expansion, which relates the product of two scaling operators to a series of local operators. Generally, when two local operators come close to each other, singularities may occur and could be encapsulated in the form $A(x)B(y) \sim  \sum C_i (x-y) O_i(y)$, where $O_i$ is a complete set of local operators. For 2D conformal invariant theories, the singular behaviour could be explicitly captured through OPE coefficients $C_{ijk}$ through
\begin{equation}
\phi_i(z,\bar{z}) \phi_j(w,\bar{w})\sim \sum_k C_{ijk} \cdot (z-w)^{h_k-h_i-h_j}  (\bar{z}-\bar{w})^{\bar{h}_k-\bar{h}_i-\bar{h}_j} \phi_k (w,\bar{w})
\end{equation}
Combing the above two equations, we could determine three point correlators as
\begin{equation}
\begin{aligned}
_{_L}\langle \phi_1(z_1,\bar{z}_1) \phi_2 (z_2,\bar{z}_2)\phi_3(z_3,\bar{z}_3)\rangle_{_R} = \frac{C_{123}}{z_{12}^{h_1+h_2-h_3} z_{23}^{h_2+h_3-h_1} z_{31}^{h_3+h_1-h_2} \cdot \bar{z}_{12}^{\bar{h}_1+\bar{h}_2-\bar{h}_3} \bar{z}_{23}^{\bar{h}_2+\bar{h}_3-\bar{h}_1}  \bar{z}_{31}^{\bar{h}_3+\bar{h}_1-\bar{h}_2} }
\end{aligned}
\end{equation}
It turns out that higher point correlation functions could also be determined by their OPE coefficients $C_{\alpha \beta \gamma}$. Again, in the general case of complex CFT, the OPE coefficients $C_{\alpha \beta \gamma}$ are complex values. 

Scaling operators could be organized into conformal towers identified through different primary operators. Within each conformal family, the descendent fields are related to the primary through Virasoro generators $L_n$ and $\bar{L}_n$. They are generators of 2D conformal transformations and their commutation relations give Virasoro algebra
\begin{equation}
\begin{aligned}
\label{eq:virasoro}
[L_n,L_m] &= (n-m)L_{n+m}+\frac{c}{12}n(n^2-1)\delta_{n+m,0}
\\
[\bar{L}_n,\bar{L}_m] &= (n-m)\bar{L}_{n+m}+\frac{c}{12}n(n^2-1)\delta_{n+m,0}
\\
[L_n,\bar{L}_m] &= 0
\end{aligned}
\end{equation}

Different from unitary CFT, in complex CFT the left CFT state is not complex conjugate of right CFT state $_L\langle \phi| \neq (|\phi\rangle_R)^{\dagger}$. In what follows, we use subscripts $L$ and $R$ to distinguish left and right eigenstates satisfying biorthogonal relations. Next we briefly discuss the properties of these complex CFT states.

For a 2D CFT, it is convenient to work on an infinite cylinder with spatial direction compactified into a circle $\mathbb{S}^1$, since this geometry could be conformally mapped to the complex plane.  Then the Hamiltonian and momentum operators generate time and space translations on the cylinder corresponding to dilation and rotation operators in the complex plane:
\begin{equation}
\begin{aligned}
H = & \frac{2\pi}{L} (L_0+\bar{L}_0-\frac{c}{12})
\\
P = & \frac{2\pi}{L}(L_0-\bar{L}_0)
\end{aligned}
\end{equation}
Employing radial quantization, the left and right eigenstates on the cylinder have one-to-one correspondence with CFT scaling operators: 
\begin{align}
|\phi\rangle_{_R} = \lim_{r\to 0}\hat{\phi} (r,0) |0\rangle_{_R} \quad _{_L}\langle\phi | = \lim_{r\to \infty} r^{2\Delta_{\phi}} ~ _{_L}\langle 0 | \hat{\phi}(r,0) 
\end{align}
where the vacuum state is defined through
\begin{equation}
\begin{aligned}
L_n|0\rangle_{_R} & = \bar{L}_n|0\rangle_{_R}=0 \quad (n\geq -1)
\\
L^{\dagger}_n|0\rangle_{_L} & = \bar{L}^{\dagger}_n|0\rangle_{_L}=0 \quad (n\geq -1).
\end{aligned}
\end{equation}
 Similar relation exists for excited states:
\begin{equation}
\begin{aligned}
L_0|\phi\rangle_{_R}  = h|\phi\rangle_{_R}, & \quad \bar{L}_0|\phi\rangle_{_R}=\bar{h} |\phi\rangle_{_R} 
\\
L^{\dagger}_0|\phi\rangle_{_L}  =h^*|\phi\rangle_{_L}, & \quad \bar{L}^{\dagger}_0|\phi\rangle_{_L}=\bar{h}^*|\phi\rangle_{_L}
\\
L_n|\phi\rangle_{_R}  =& \bar{L}_n|\phi\rangle_{_R}=0 \quad (n > 0)
\\
L^{\dagger}_n|\phi\rangle_{_L}  =& \bar{L}^{\dagger}_n|\phi\rangle_{_L}=0 \quad (n>0)
\end{aligned}
\end{equation}

This state-operator correspondence is crucial to extract conformal data within lattice computation. The Virasoro generators could be used to generate states with higher conformal dimensions. For example, for a primary state $|\phi\rangle_{_R}$ ($|\phi\rangle_{_L}$), we could obtain its descendent states by applying $L_n$ ($L^{\dagger}_n$) successively
\begin{equation}
\begin{aligned}
L_{-k_1}L_{-k_2} \cdots L_{-k_m} |\phi\rangle_{_R}, & \quad (1\leq k_1 \leq k_2 \cdots \leq k_m)
\\
L^{\dagger}_{-k_1}L^{\dagger}_{-k_2} \cdots L^{\dagger}_{-k_m} |\phi\rangle_{_L}, & \quad (1\leq k_1 \leq k_2 \cdots \leq k_m),
\end{aligned}
\end{equation}
which should have conformal weight $h'=h+\sum k_i$ and $\bar{h}' = \bar{h}$. The anti-holomorphic partners $\bar{L}_n$ ($\bar{L}^{\dagger}_n$) could be used to increase $\bar{h}$ similarly. Each primary state together with its descendents span a closed Hilbert space and form a representation of the underlying Virasoro algebra. Each individual subspace is called Verma module as illustrated in Tab~\ref{tab:vm}.

\begin{table}
\caption{\label{tab:vm}Verma module for state $|\phi\rangle_{_R}$ up to level 4}
\begin{tabular}{ccl}
\hline
 level & $h$ & state\\  \hline

0 & $h$ & $|\phi\rangle_{_R}$  \\ 

1 & $h+1$ & $L_{-1}|\phi\rangle_{_R}$  \\ 

2 & $h+2$ & $L^2_{-1}|\phi\rangle_{_R},L_{-2}|\phi\rangle_{_R}$   \\ 

3 & $h+3$ & $L^3_{-1}|\phi\rangle_{_R},L_{-1}L_{-2}|\phi\rangle_{_R},L_{-3}|\phi\rangle_{_R}$   \\ 

4 & $h+4$ & $L^4_{-1}|\phi\rangle_{_R},L^2_{-1}L_{-2}|\phi\rangle_{_R},L^2_{-2}|\phi\rangle_{_R},L_{-1}L_{-3}|\phi\rangle_{_R},L_{-4}|\phi\rangle_{_R}$   \\ 

 && $\cdots$ \\
\hline

\end{tabular}
\end{table}

Crucially, not all states within the Verma module could be identified within the energy spectrum, since the linear combination of some states (called null states) may vanish. For instance, considering level 1, we get $L_{-1}|\phi\rangle_{_R} = 0$, this corresponds to $h=0$ and $|\phi\rangle_{_R}=|0\rangle_{_R}$. Thus the first level descendent of vaccum state is always null state. Considering level 2, we get $L_{-2}|\phi\rangle_{_R}+a L^2_{-1}|\phi\rangle_{_R}= 0$. Solving this equation, one get $c=2h(5-8h)/(2h+1)$.
These features of Virasoro algebra have been explicitly identified in our model calculations. For example, we verify the conformal weight of leading singlet operator $h_{\epsilon}$ fulfills this condition and its level-2 descendent states are indeed only one-fold degenerate as expected from null field condition (see Fig. 3 in the main text). Similar analysis suggests the third-order descendent of $ |\epsilon'\rangle_{_R}$ in our model is also a null state satisfying $L_{-3} |\epsilon'\rangle_{_R} + aL_{-1}L_{-2}|\epsilon'\rangle_{_R}+b L^3_{-1}|\epsilon'\rangle_{_R}= 0$.

\section{B. Location of phase transition point from the conformal perturbation calculation}

The phase transition point of the microscopic lattice Hamiltonian $H_{\text{NH-Potts}}$ can be determined as following. 
Around the CFT fixed point, we take the effective Hamiltonian to be a complex fixed point Hamiltonian with perturbations:
\begin{align}
H_{\text{NH-Potts}} (J=1,h=1,\lambda) \propto H_{\text{CFT}}(g_{\epsilon'})= H_{\text{CFP}} + g_{\epsilon'} \int \mathrm{d}x \cdot \epsilon'(x) + ... 
\end{align}
where $H_{\text{CFP}}$ is the exact Hamiltonian at the complex fixed point, whose eigenenergy spectrum is precisely the operator spectrum of the complex CFT. Since we only study the Hamiltonian with $S_5$ symmetry and Kramers-Wannier duality, the only symmetry allowed relevant operator is $\epsilon'$. So the CFT fixed point is realized when $g_{\epsilon'}=0$ in the thermodynamic limit. Next we treat an effective coupling $g_{\epsilon'}$ as a free parameter and the minimization of $g_{\epsilon'}$ should correspond to the CFT fixed point $H_{\text{lattice}} (J=1,h=1,\lambda=\lambda_c) \propto H_{\text{CFT}}(0)= H_{\text{CFP}} $.

In specific, we assume that the coupling strength $g_{\epsilon'}$ is
small and apply the first-order  perturbation theory \cite{P_Reinicke_1987a,P_Reinicke_1987b,Lao2023}:
\begin{align}
\delta E_\phi = \frac{ _{_L}\langle \phi|\delta H|\phi \rangle _{_R}}{ _{_L}\langle\phi |\phi \rangle _{_R} }
\end{align}
Then for a scalar primary $\phi$, we will get 
\begin{align}
\delta E_\phi &= g_{\epsilon'} C_{\phi \epsilon'\phi},
\\
\delta E_{\partial\phi} &= g_{\epsilon'} C_{\phi \epsilon'\phi} A_{\partial \phi,\epsilon'}, \quad  A_{\partial \phi,\epsilon'} = 1+\frac{\Delta_{\epsilon'}(\Delta_{\epsilon'}-2)}{4\Delta_{\phi}}
\end{align}
in terms of the OPE coefficients $C_{\phi\epsilon' \phi}$.  
Thus the energy gap for the perturbed Hamiltonian reads
\begin{equation}
    \alpha(E_\phi-E_0) \approx \Delta_\phi+\delta E_\phi
\end{equation}
up to a finite-size correction from higher primary/descendent fields. Next, one could minimize the following cost function to evaluate $\alpha$ and $g_{\epsilon'}$ for a given $\lambda$
\begin{equation}
    \delta(\alpha,g_{\epsilon'}) = \sum_n |\alpha (E_n-E_0)-\Delta_n-\delta E_n|^2.
\end{equation}
The position with minimal $g_{\epsilon'}$  corresponds to the phase transition point $\lambda=\lambda_c$ \cite{Lao2023}. Here, we numerically test this scheme for 4 states corresponding to $\sigma$, $\partial \sigma$, $\epsilon$ and $\partial \epsilon$. For $H_{\text{NH-Potts}}$ at different total system size $L=8-12$, the critical point $(J_c,h_c,\lambda_c)$ is determined by the location of minimum of coupling strength $g_{\epsilon'}$, as shown in Fig~\ref{fig:minimize_g}.

Although the discussion within this work concentrated on $Q=5$ state Potts model, this procedure is also applicable to locate the complex fixed points for other integer $Q \geq 6$ within our model $H_{\text{NH-Potts}}(J,h,\lambda)$.

\begin{figure*}
\includegraphics[width=0.35\textwidth]{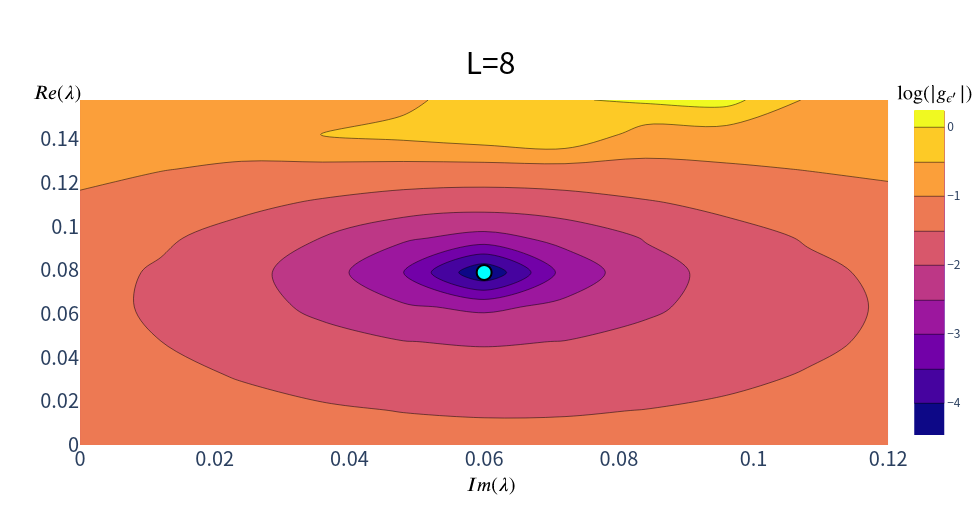}
\includegraphics[width=0.35\textwidth]{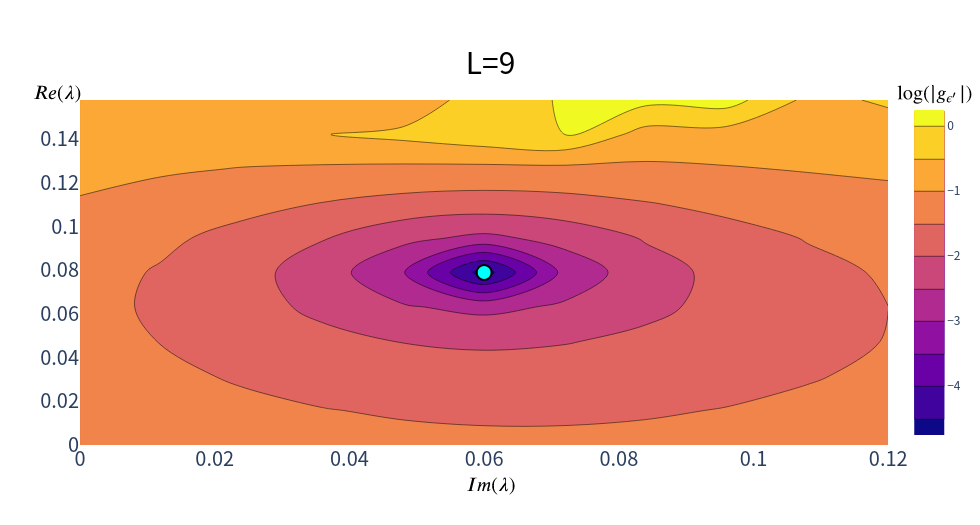}
\includegraphics[width=0.35\textwidth]{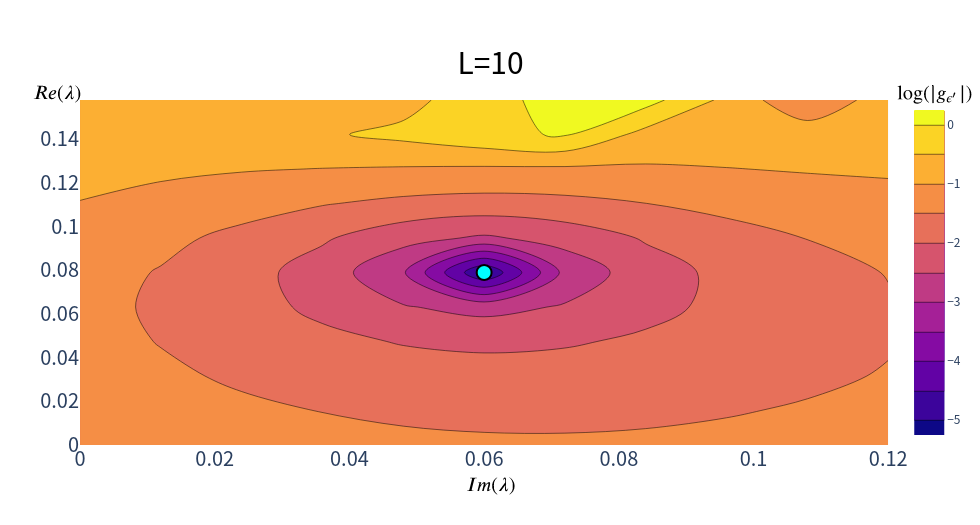}
\includegraphics[width=0.35\textwidth]{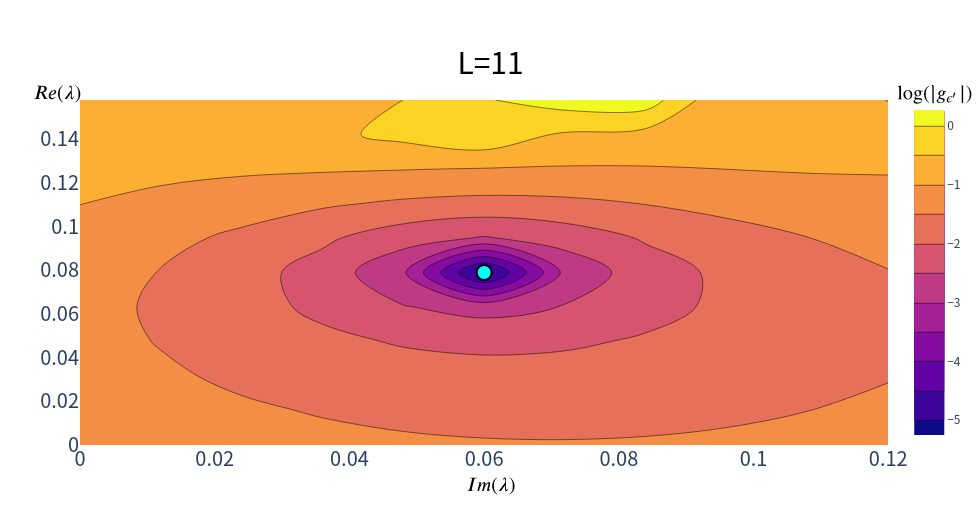}
\includegraphics[width=0.35\textwidth]{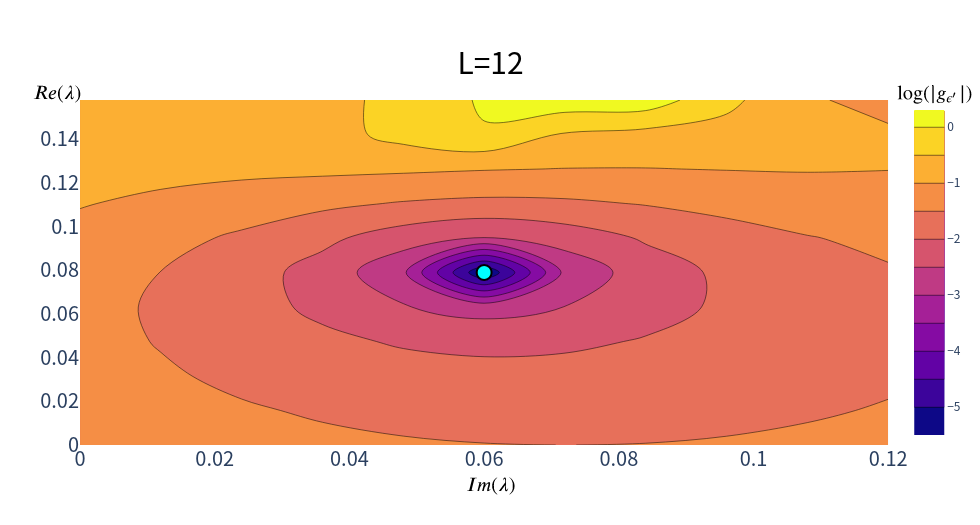}
\caption{\label{fig:minimize_g} The distribution of optimized coupling strength $g_{\epsilon'}$ by scanning $\lambda$ on the self-duality plane $J=h=1$. The subfigures are for system size $L=8,9,10,11,12$. The minimum of $|g_{\epsilon'}|$ is shown by the cyan dot, which gives the critical point $(J_c=1,h=1,\lambda_c=0.079+0.060i)$ of microscopic Hamiltonian $H_{\text{NH-Potts}}$.  }
\end{figure*}

\section{C. Finite-size correction to scaling dimensions}

By sitting at the critical point, the critical lattice Hamiltonian may deviate from the exact criticality due to the corrections from irrelevant operators \cite{CARDY1986}. Considering the constraint from symmetry and duality, our critical lattice model could be approximated by
\begin{equation}
\begin{aligned}
\label{eq:irrelevant_correction}
H_{\text{NH-Potts}} (1,1,\lambda_c) \propto  H_{\text{CFP}} + a_{\Box \epsilon'} \int \mathrm{d}v\cdot \Box \epsilon'(v)
 + a_{T\overline{T}} \int \mathrm{d}v\cdot T\overline{T}(v) + a_{T^2+\overline{T}^2} \int \mathrm{d}v\cdot (T^2+\overline{T}^2)(v) +\cdots
\end{aligned}
\end{equation}  
where $\Box \epsilon'$, $T\overline{T}$ and $T^2+\overline{T}^2$ are three leading irrelevant operators respecting same symmetry as the original model with $a_{\Box \epsilon'}$ and $a_{T\overline{T}}$ to be some unknown coefficients. $H_{\text{CFP}}$ corresponds to the exact complex fixed point defined on $S^1 \times R^1$ with its eigenenergy spectrum forming representation of underlying conformal group.

Using first-order perturbation theory, the energy gap for the model \eqref{eq:irrelevant_correction} can be expressed as
\begin{equation}
\begin{aligned}
E_n-E_0 \propto  \frac{2\pi \Delta_n}{L} [ 1 + (a_{T\overline{T}} C_{nn,T\overline{T}} + a_{T^2+\overline{T}^2} C_{nn,T^2+\overline{T}^2})(\frac{2\pi}{L})^2 + \cdots ]
\end{aligned}
\end{equation}
where $C_{nn,T\overline{T}}$ and $C_{nn,T^2+\overline{T}^2}$ are OPE coefficients. The $1/L^2$ contribution comes from $\Delta_{T\overline{T}}=\Delta_{T^2}=\Delta_{\overline{T}^2}=4$, which exists for any two-dimensional CFT. To verify the exact complex fixed points, we fit the energy gap $(E_n-E_0)$ for some low-lying states with fitting formula $y=A+\frac{B}{L}+\frac{C}{L^3}$. In contrast to unitary CFTs, here all fitting coefficients ($A,B$ and $C$) are complex. At the complex fixed point, the band gap should be scaled to $0$, which means $A\approx0+0i$. 
The scaling of the band gap at this critical point is illustrated in Fig.2 of the main text.

To obtain the operator dimension, we normalize the eigenenergy corresponds to the energy-momentum tensor to be $\Delta_T = 2+0i$. This is achieved by numerically evaluating
\begin{equation}
\begin{aligned}
\label{eq:ope_dim_scaling}
\frac{2(E_n-E_0)}{E_T-E_0} =  2 \cdot \frac{ \frac{2\pi \xi \Delta_n}{L} + \frac{\beta_n}{L^3} +\cdots}{\frac{2\pi \xi \Delta_T}{L} + \frac{\beta_T}{L^3} +\cdots} \approx  \Delta_n  + \frac{\beta_{nT}}{L^2}+\cdots
\end{aligned}
\end{equation}
To get a more accurate estimation of $\Delta_n$, we firstly calculate the LHS of \eqref{eq:ope_dim_scaling} at different lattice size $L$ and extrapolate the result to thermodynamic limit $L\to\infty$. The scaling of the real and imaginary parts of 12 (quasi-)primary operators are illustrated in Fig~\ref{fig:scale_ope_dim}. The detailed conformal multiplets for different primary operators are listed from Tab~\ref{tab:con_familiy_epsilon} to Tab~\ref{tab:con_familiy_T}.

\graphicspath{{sup_fig/}}
\begin{figure*}
\includegraphics[width=0.95\textwidth]{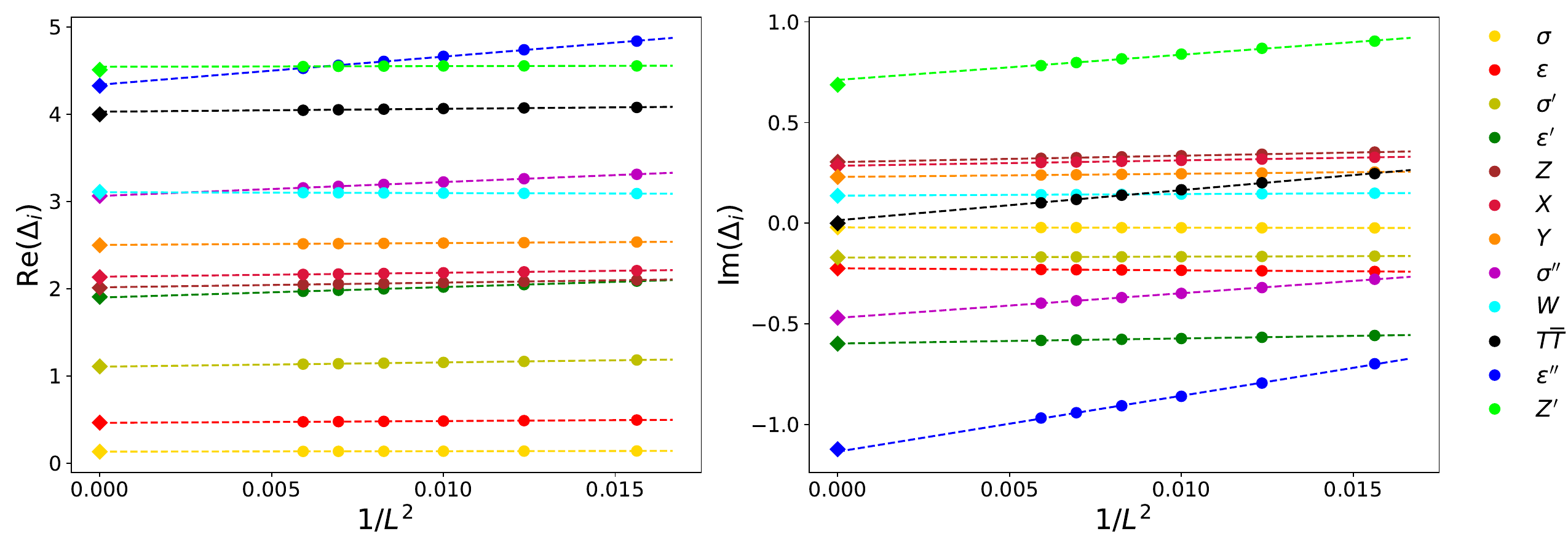}
\caption{\label{fig:scale_ope_dim} Scaling of the real (left) and imaginary (right) part of the operator dimensions. The round circles are numerical results evaluated at different total system size $L$ with different operators distinguished by colors. The dashed lines are fitted from \eqref{eq:ope_dim_scaling}. The diamonds denote prediction from complex CFT \cite{Gorbenko2018b}. }
\end{figure*}

\section{D. Central charge}

For one-dimensional critical system with periodic boundary condition, the central charge $c$ can be extracted through the ground state free energy density \cite{Affleck1988,Blote1986}
\begin{equation}
\begin{aligned}
\label{eq:fit_c}
\frac{E_0}{L} = E_g - \frac{\pi c v}{6L^2} +\cdots
\end{aligned}
\end{equation}
where $c$ is the central charge and $v$ is a model dependent constant. To determine the coefficient $v$, we use the relationship
\begin{equation}
\begin{aligned}
E_n-E_0 =  \frac{2\pi v \Delta_n}{L} + \frac{\beta_n}{L^3} +\cdots,
\end{aligned}
\end{equation}
and we choose to fit the state for energy-momentum tensor with $\Delta_T=2$ and get $v \approx 2.8810-0.7091i$. Then we fit the ground state free energy density via \eqref{eq:fit_c} and get $c\approx 1.1405-0.0224i$, as shown in Fig.~\ref{fig:scale_c}. This is close to the theoretical prediction from the analytical continuation $c_{\text{5-potts}} \approx 1.1375-0.0211i$ \cite{Gorbenko2018b}.

\begin{figure*}
\includegraphics[width=0.8\textwidth]{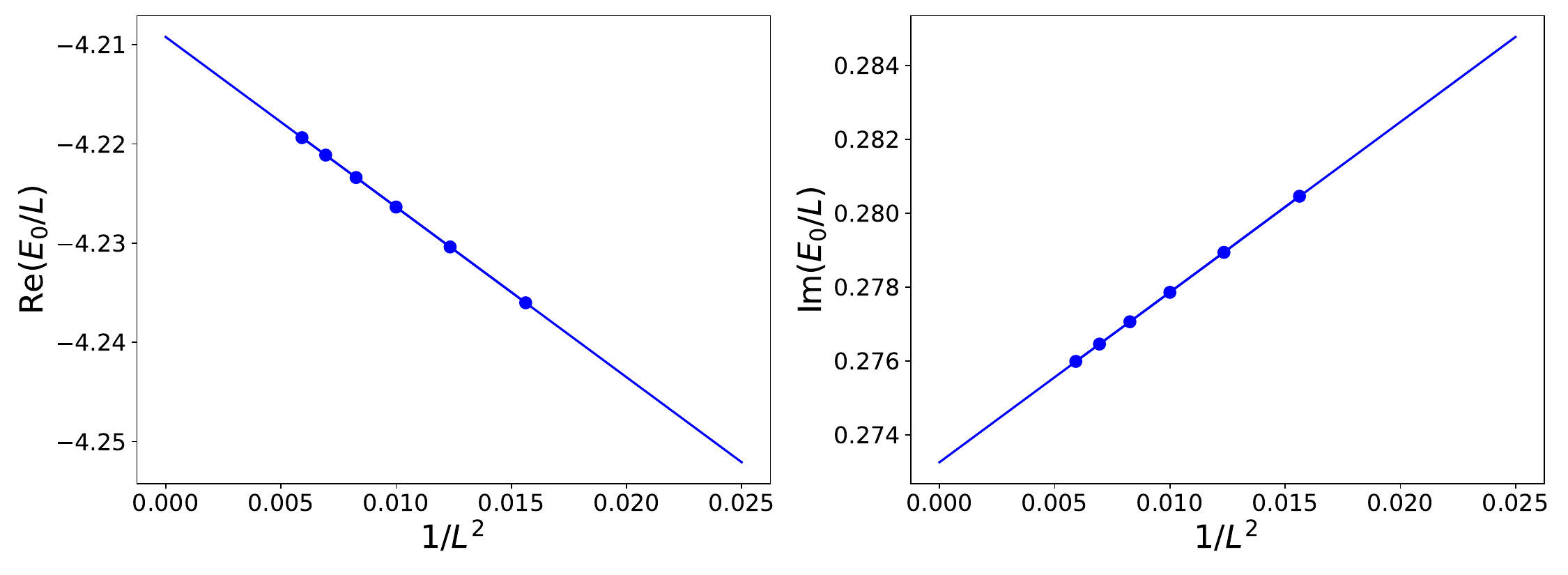}
\caption{\label{fig:scale_c} Scaling of the real and imaginary part of ground state free energy per unit length $E_0/L$, which gives rise to the central charge $c\approx 1.1405-0.0224i$. 
}
\end{figure*}

\section{E. Operator spectra of original 5-state Potts model}
In the main text, we have shown the operator spectra of 5-state Potts model at the complex critical point $(J_c=1,h_c=1,\lambda_c)$. As a comparison, here we show the energy spectra at the weakly first-order transition point $(J=1,h=1,\lambda=0)$ in the original 5-state Potts model, as shown in Fig. \ref{fig:wfo_conformal_familiy}. 
The main finding is, the spectra of weakly first-order transition develops an approximate conformal tower structure, i.e. the energy levels largely fit, but some of levels clearly deviate from the  CFT expectations.  One plausible reason is, the weakly first-order transition point is very close to the complex fixed point (see Fig. 1 and 2 in the main text), so its spectra just inherits the conformal structure of CFT fixed point.

\begin{figure*}[!htb]
\includegraphics[width=0.8\textwidth]{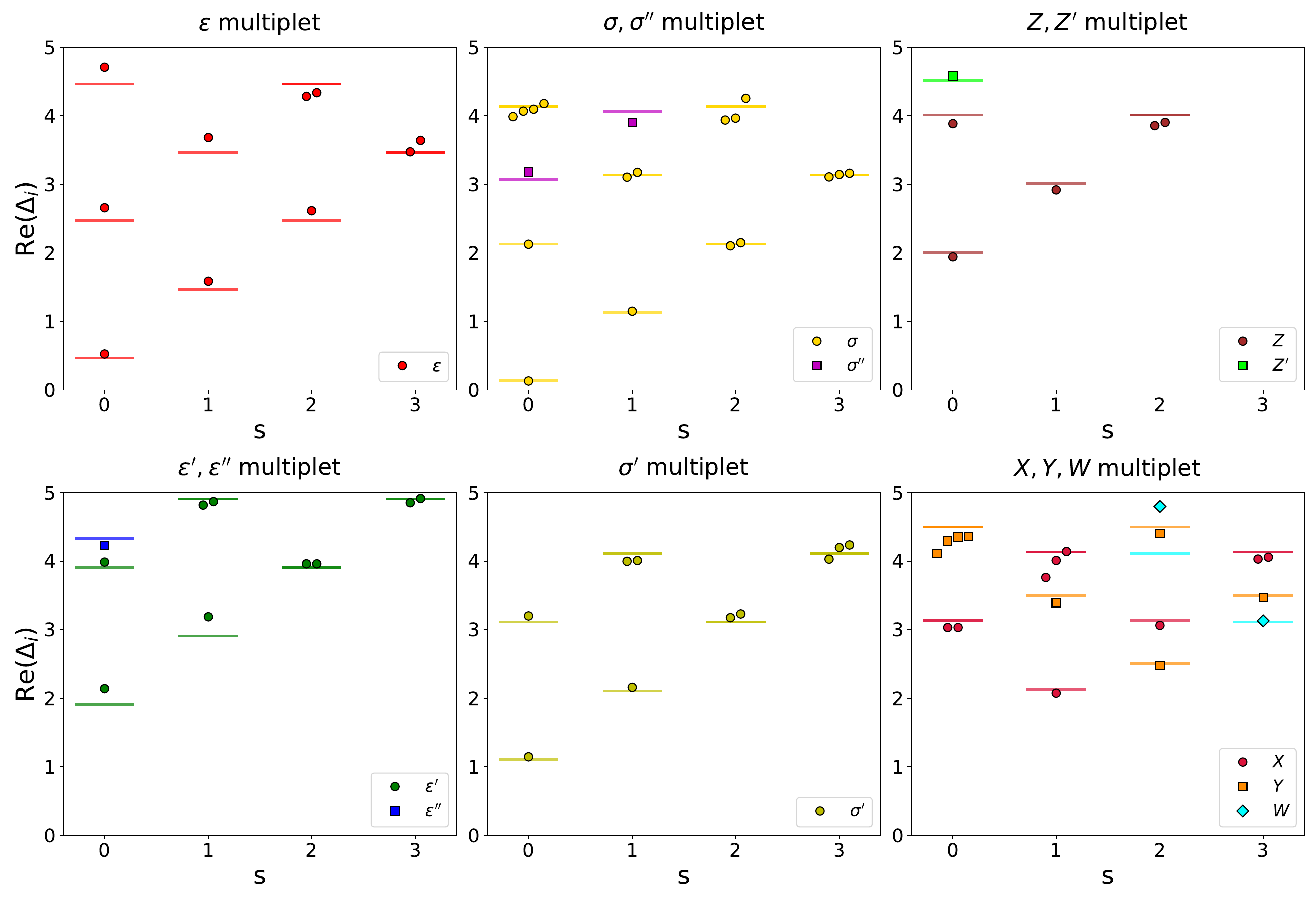}
\caption{\label{fig:wfo_conformal_familiy} Energy spectra of original 5-state Potts model at the weakly first-order transition point $(J=1,h=1,\lambda=0)$. The results show an approximate tower structure, where many descendent fields deviate away from the expectation values but not far away. }
\end{figure*}

\section{F. Operator content of local operators}

In principal, all local lattice operators $O$ can be expanded into linear combination of CFT operators $\hat \phi_{\alpha}$
\begin{equation}
O=\sum_{\alpha} c_{\alpha} \hat{\phi}_{\alpha},
\end{equation}
where the summation is generally over infinite primary and descendant operators. The coefficients in front of  CFT operators respecting different symmetry of the lattice operator is guaranteed to vanish. For example, the lattice operator $\sigma_i$ has charge $1$ under $Z_Q$ spin cyclic symmetry. Consequently, it cannot have any expansion over $Z_Q$ charge $0$ CFT operators such as $\epsilon$, $\epsilon'$, $\epsilon''$ or their descendents.  From the perspective of numerical calculation, the non-universal coefficient $c_{\alpha}$ is zero if and only if $_{_L}\langle \phi_{\alpha}|O|0\rangle_{_R} = ~ _{_L}\langle 0|O|\phi_{\alpha}\rangle_{_R}=0$ (for scalar operator $O$). Based on the symmetry constraint, we write down the operator decomposition as

\begin{equation}\label{eq:content_sigma}
\begin{aligned}
O_{\sigma} = & \sigma_i = (a_{\sigma} \cdot \hat{\sigma} +\text{desc.})
 +(b_{\sigma'} \cdot \hat{\sigma}' + \text{desc.}) + \cdots
\end{aligned}
\end{equation}

\begin{equation}\label{eq:content_epsilon}
\begin{aligned}
 O_{\epsilon} = & \sum_{k=1}^{Q-1} [ (\sigma_i^{\dagger} \sigma_{i+1})^k - \tau_i^k    ]  
=  (a_{\epsilon} \cdot \hat{\epsilon} + \text{desc.}) 
+ (a_{\epsilon''} \cdot \hat{\epsilon}'' +  \text{desc.} ) + \cdots
\end{aligned}
\end{equation}

\begin{equation}\label{eq:content_epsilon'}
\begin{aligned}
O_{\epsilon'1} & = \sum_{k_1=1}^{Q-1} \sum_{k_2=1}^{Q-1} [(\tau_i^{k_1} +\tau_{i+1}^{k_1}) (\sigma_i^{\dagger} \sigma_{i+1})^{k_2}
 + (\sigma_i^{\dagger} \sigma_{i+1})^{k_1} (\tau_i^{k_2} +\tau_{i+1}^{k_2}) ]
\\
&=  a_{ \mathbb{I}} \cdot  \hat{\mathbb{I}} + (b_{\epsilon'} \cdot \hat{\epsilon}' +  \text{desc.}) +c_{T,\overline{T}} \cdot (\hat{T}+\hat{\overline{T}}) + c_{T\overline{T}} \cdot \hat{T\overline{T}} + \cdots
\end{aligned}
\end{equation}
The dots represents other CFT operators with higher scaling dimensions. We may also verify these expansion through local correlators of primary operators.
Firstly, we recall that the two-point correlator of primary operator $\hat{\phi}$ on a complex plane ($z,\bar{z}=x\pm it$) is:
\begin{equation}
    _{_L}\langle 0 | \hat{\phi}(x_1,0) \hat{\phi}(x_2,0) \rangle_{_R} = \frac{1}{(x_1-x_2)^{2\Delta_{\phi}}}
\end{equation}
where $\Delta_{\phi}$ is its scaling dimension. Employing the state-operator correspondence
\begin{equation}
    |\phi\rangle_{R} = \lim_{r\to 0}\hat{\phi} (r,0) |0\rangle_{R}, \quad {_L}\langle\phi | = \lim_{r\to \infty} r^{2\Delta_{\phi}} ~ {_L}\langle 0 | \hat{\phi}(r,0) 
\end{equation}
we have 
\begin{equation}
\begin{aligned}
     _L\langle 0 | \hat{\phi}(x,0) | \phi \rangle_R = \lim_{x'\to 0 } {_L}\langle 0 | \hat{\phi}(x,0) \hat{\phi}(x',0) | 0 \rangle_R 
     = \frac{1}{x^{2\Delta_{\phi}}}
     \\
     _L\langle \phi | \hat{\phi}(x,0) | 0 \rangle_R = \lim_{x'\to \infty } x'^{2\Delta_{\phi}} {_L}\langle 0 | \hat{\phi}(x',0) \hat{\phi}(x,0) | 0 \rangle_R 
     = 1
\end{aligned}
\end{equation}
Next, we apply a Weyl transformation $w=R\ln z$ to map the complex plane into an infinite cylinder $\mathbb{S}^1 \times \mathbb{R}^1$. The operator transform accordingly as
\begin{equation}
    \hat{\phi}(w) =  \big{(}\frac{\mathrm{d}\omega}{\mathrm{d}z} \big{)}^{-\Delta_{\phi}} \hat{\phi}(z)=  \big{(}\frac{R}{z} \big{)}^{-\Delta_{\phi}} \hat{\phi}(z)
\end{equation}
Then one could evaluate the local correlator within the cylinder from
\begin{equation}
    {_L}\langle 0 | \hat{\phi}(w) | \phi \rangle_{R} {_L}\langle \phi | \hat{\phi}(w) | 0 \rangle_{R} =~  {_L}\langle 0 | \big{(}\frac{R}{z} \big{)}^{-\Delta_{\phi}} \hat{\phi}(z) | \phi \rangle_{R}  {_L}\langle \phi | \big{(}\frac{R}{z} \big{)}^{-\Delta_{\phi}} \hat{\phi}(z) | 0 \rangle_{R} 
    = R^{-2\Delta_{\phi}}
\end{equation}

When we calculate the correlators using the  lattice operators, we can derive the expansion form as following, by using the CFT operators 

\begin{equation}
\begin{aligned}
\sqrt{_{_L}\langle 0| O_{\sigma}|\sigma\rangle_{_R} {}_{_L}\langle \sigma| O_{\sigma}|0\rangle_{_R} } = 
& (a_{\sigma} (\frac{2\pi}{L})^{\Delta_{\sigma}}+a_{\Box \sigma} (\frac{2\pi}{L})^{\Delta_{\sigma}+2}+\cdots)
\\
& +(b_{\sigma'} (\frac{2\pi}{L})^{\Delta_{\sigma'}}+b_{\Box \sigma'} (\frac{2\pi}{L})^{\Delta_{\sigma'}+2}+\cdots) 
\\
& +\cdots
\end{aligned}
\end{equation}

\begin{equation}
\begin{aligned}
\sqrt{_{_L}\langle 0| O_{\epsilon}|\epsilon\rangle_{_R} {}_{_L}\langle \epsilon| O_{\epsilon}|0\rangle_{_R} } = 
& (a_{\epsilon} (\frac{2\pi}{L})^{\Delta_{\epsilon}}+a_{\Box \epsilon} (\frac{2\pi}{L})^{\Delta_{\epsilon}+2}+\cdots)
\\
& + (b_{\epsilon''} (\frac{2\pi}{L})^{\Delta_{\epsilon''}}+\cdots)  +\cdots
\end{aligned}
\end{equation}

\begin{equation}
\begin{aligned}
\sqrt{_{_L}\langle 0| O_{\epsilon'1}|\epsilon'\rangle_{_R} {}_{_L}\langle \epsilon'| O_{\epsilon'1}| 0 \rangle_{_R} } = 
& (b_{\epsilon'} (\frac{2\pi}{L})^{\Delta_{\epsilon'}}+b_{\Box \epsilon'} (\frac{2\pi}{L})^{\Delta_{\epsilon'}+2}+\cdots)
\\
& + (c_{T\overline{T}} (\frac{2\pi}{L})^{\Delta_{T\overline{T}}}+\cdots) +\cdots
\end{aligned}
\end{equation}

Since biorthogonal eigenstates are invariant under an overall rescaling $|\psi\rangle_{_R} \to \gamma |\psi\rangle_{_R} $ and $|\psi\rangle_{_L} \to (\gamma^*)^{-1} |\psi\rangle_{_L} $, where $\gamma$ is a non-zero complex constant, we combine both left and right eigenstates to get rid of this arbitrariness. Then we can fit the two-point function using Eq. \ref{eq:fit_sigma}, \ref{eq:fit_epsilon}, \ref{eq:fit_epsilon'} to verify the operator content. The numerical fitting curves are illustrated in Fig~\ref{fig:scale_delta}. The agreement with the fitting curve demonstrates that the content of the lattice operators indeed satisfy Eq. \ref{eq:content_sigma} \ref{eq:content_epsilon}, \ref{eq:content_epsilon'}. 

\begin{figure*}
\includegraphics[width=0.3\textwidth]{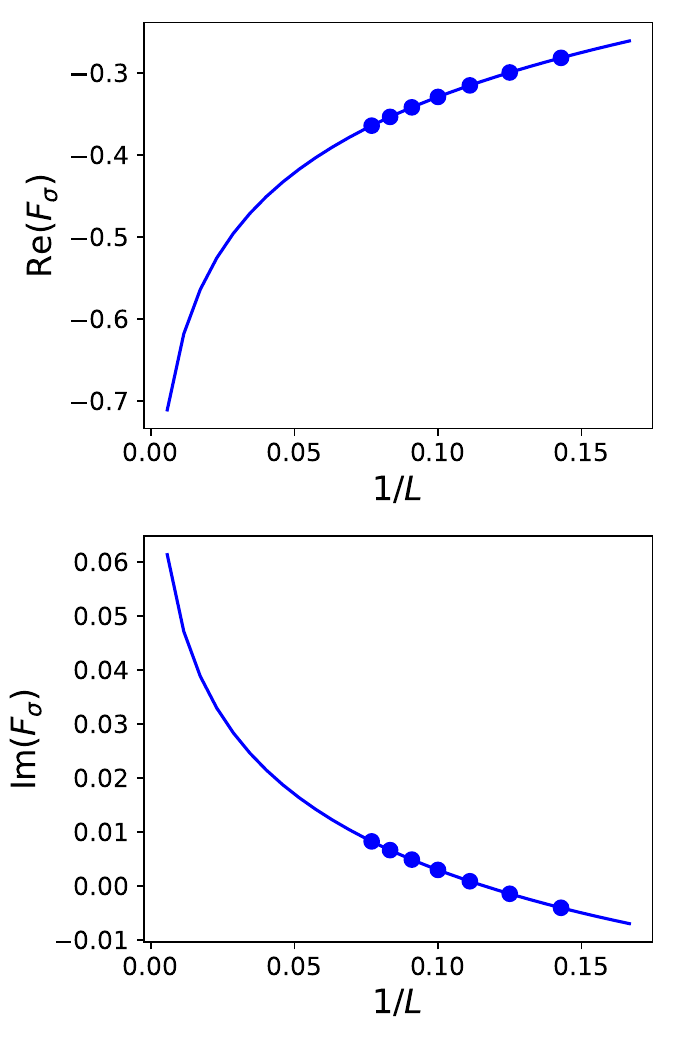}
\includegraphics[width=0.3\textwidth]{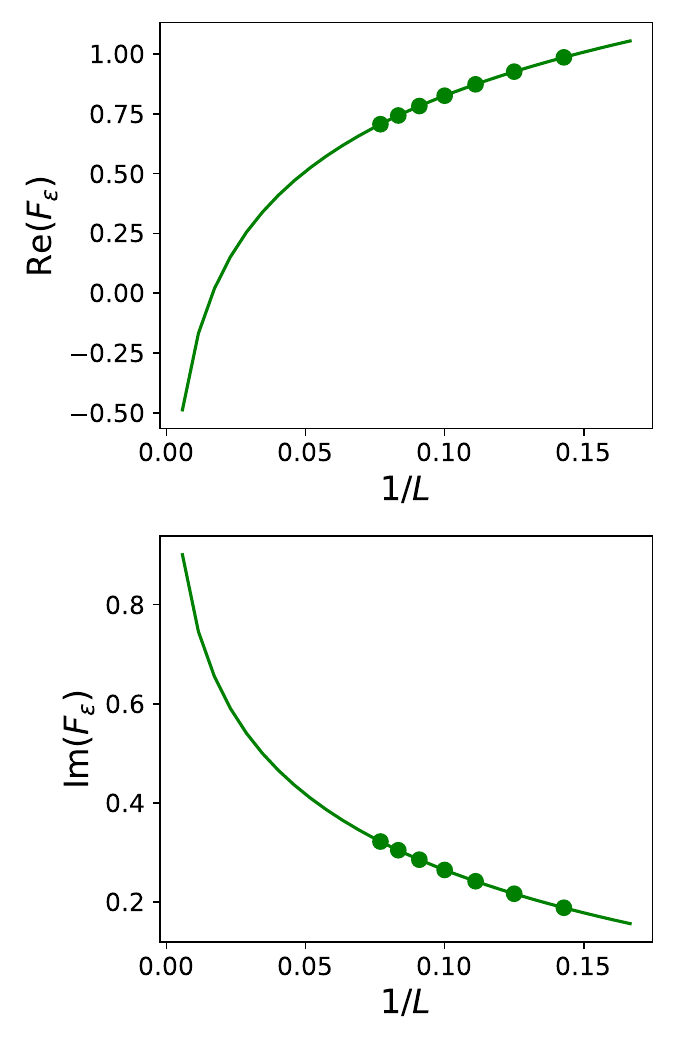}
\includegraphics[width=0.3\textwidth]{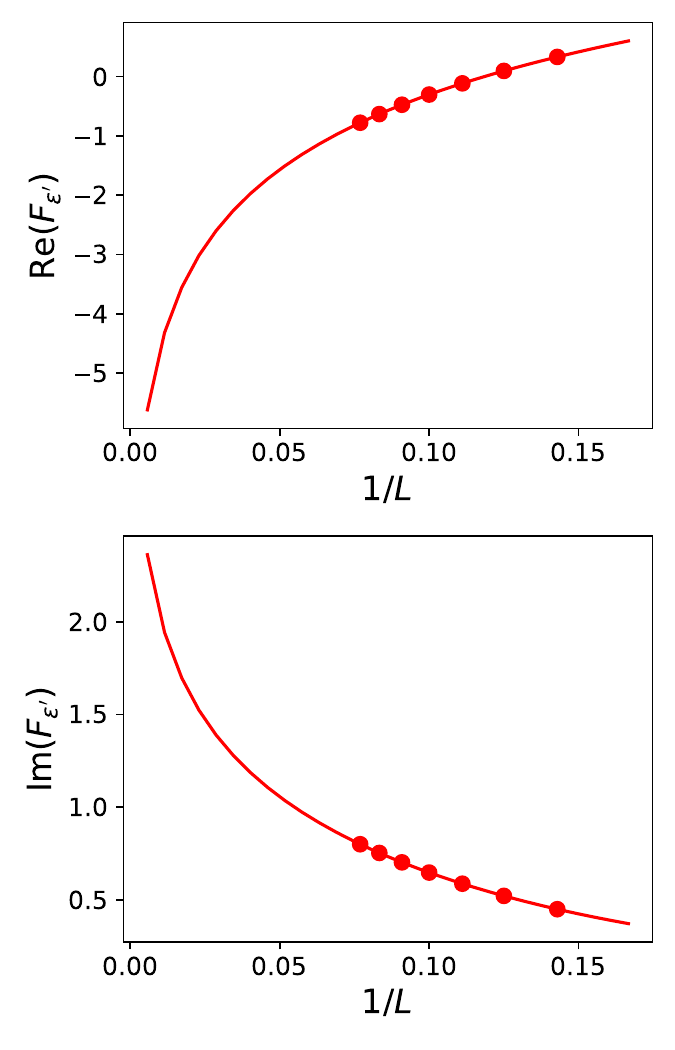}
\caption{\label{fig:scale_delta} Finite-size scaling of the two-point functions using Eq. \ref{eq:fit_sigma}, \ref{eq:fit_epsilon}, \ref{eq:fit_epsilon'}.  }
\end{figure*}

\begin{equation}
\begin{aligned}
\label{eq:fit_sigma}
F_{\sigma} = \ln \sqrt{_{_L}\langle 0| O_{\sigma}|\sigma\rangle_{_R} {}_{_L}\langle \sigma| O_{\sigma}|0\rangle_{_R} } \approx
 A_{\sigma} - \Delta_{\sigma} \ln L + \frac{B_{\sigma}}{L^{\Delta_{\sigma'}-\Delta_{\sigma}}} + \frac{C_{\sigma}}{L^2}  + o(\frac{1}{L^{\Delta_{\sigma'}+2-\Delta_{\sigma}}})
\end{aligned}
\end{equation}

\begin{equation}
\begin{aligned}
\label{eq:fit_epsilon}
F_{\epsilon} = \ln \sqrt{_{_L}\langle 0| O_{\epsilon}|\epsilon\rangle_{_R} {}_{_L}\langle \epsilon| O_{\epsilon}|0\rangle_{_R} } \approx
 A_{\epsilon} - \Delta_{\epsilon} \ln L + \frac{B_{\epsilon}}{L^2} + \frac{C_{\epsilon}}{L^{\Delta_{\epsilon''}-\Delta_{\epsilon}}}  + o(\frac{1}{L^4})
\end{aligned}
\end{equation}

\begin{equation}
\begin{aligned}
\label{eq:fit_epsilon'}
F_{\epsilon'} = \ln \sqrt{_{_L}\langle 0| O_{\epsilon'1}|\epsilon'\rangle_{_R} {}_{_L}\langle \epsilon'| O_{\epsilon'1}|0\rangle_{_R} } \approx
 A_{\epsilon'} - \Delta_{\epsilon'} \ln L + \frac{B_{\epsilon'}}{L^2} + \frac{C_{\epsilon'}}{L^{\Delta_{T\overline{T}}-\Delta_{\epsilon'}}}  + o(\frac{1}{L^4})
\end{aligned}
\end{equation}

\section{G. Finite-size correction to OPE coefficients}

The identification of operator content of lattice operators  enables us to extract the OPE coefficients. Next we briefly outline the procedure.

Employing the decomposition of $O_{\sigma}$, we can evaluate the OPE coefficient $C_{\alpha \sigma \beta}$ from the following quantity 
\begin{equation}
\begin{aligned}
\label{eq:ope_sigma}
\sqrt{ \frac{  _{_L}\langle \alpha| O_{\sigma}|\beta\rangle_{_R} {}_{_L}\langle \beta| O_{\sigma}|\alpha\rangle_{_R}  }{ _{_L}\langle 0| O_{\sigma}|\sigma\rangle_{_R} {}_{_L}\langle \sigma| O_{\sigma}|0\rangle_{_R} }  } &= 
\frac{  a_{\sigma} C_{\alpha \sigma \beta} (\frac{2\pi}{L})^{\Delta_{\sigma}}+a_{\Box \sigma} C_{\alpha ,\Box \sigma, \beta} (\frac{2\pi}{L})^{\Delta_{\sigma}+2}
 +b_{\sigma'} C_{\alpha \sigma' \beta} (\frac{2\pi}{L})^{\Delta_{\sigma'}} + \cdots  }{   a_{\sigma} (\frac{2\pi}{L})^{\Delta_{\sigma}}+a_{\Box \sigma} (\frac{2\pi}{L})^{\Delta_{\sigma}+2}
 +b_{\sigma'} (\frac{2\pi}{L})^{\Delta_{\sigma'}} + \cdots  }
\\
& \approx  C_{\alpha \sigma \beta} + \frac{c_1}{L^{\Delta_{\sigma'}-\Delta_{\sigma}}} + \frac{c_2}{L^{2(\Delta_{\sigma'}-\Delta_{\sigma})}}+ \frac{c_3}{L^2}+o(\frac{1}{L^{\Delta_{\sigma'}-\Delta_{\sigma}+2}}),
\end{aligned}
\end{equation}
where $c_{1,2,3}$ are non-universal coefficients. 
Please note that, the form of $_{_L}\langle \alpha| O_{\sigma}|\beta\rangle_{_R} {}_{_L}\langle \beta| O_{\sigma}|\alpha\rangle_{_R} $ automatically removes the arbitrary global phase of the left and right eigenstates.  
And we expect to obtain $C_{\alpha \sigma \beta}$ by finite-size extrapolation.

Similarly, the finite-size scaling for OPE coefficient $C_{\alpha \epsilon \beta}$ reads

\begin{equation}
\begin{aligned}
\label{eq:ope_epsilon}
\sqrt{ \frac{  _{_L}\langle \alpha| O_{\epsilon}|\beta\rangle_{_R} {}_{_L}\langle \beta| O_{\epsilon}|\alpha\rangle_{_R}  }{ _{_L}\langle 0| O_{\epsilon}|\epsilon\rangle_{_R} {}_{_L}\langle \epsilon| O_{\epsilon}|0\rangle_{_R} }  } & = 
\frac{  a_{\epsilon} C_{\alpha \epsilon \beta} (\frac{2\pi}{L})^{\Delta_{\epsilon}}+a_{\Box \epsilon} C_{\alpha ,\Box \epsilon, \beta} (\frac{2\pi}{L})^{\Delta_{\epsilon}+2}
 +b_{\epsilon''} C_{\alpha \epsilon'' \beta} (\frac{2\pi}{L})^{\Delta_{\epsilon''}} + \cdots  }{   a_{\epsilon} (\frac{2\pi}{L})^{\Delta_{\epsilon}}+a_{\Box \epsilon} (\frac{2\pi}{L})^{\Delta_{\epsilon}+2}
 +b_{\epsilon''} (\frac{2\pi}{L})^{\Delta_{\epsilon''}} + \cdots  }
\\
& \approx C_{\alpha \epsilon \beta} + \frac{d_1}{L^2} + \frac{d_2}{L^{{\Delta_{\epsilon''}-\Delta_{\epsilon}}}}+o(\frac{1}{L^4})
\end{aligned}
\end{equation}

Especially, the OPE coefficient $C_{\alpha \epsilon' \beta}$ for $\alpha \neq \beta$ can be obtained by
\begin{equation}
\begin{aligned}
\label{eq:ope_epsilon'}
\sqrt{ \frac{  _{_L}\langle \alpha| O_{\epsilon'}|\beta\rangle_{_R} {}_{_L}\langle \beta| O_{\epsilon'}|\alpha\rangle_{_R}  }{ _{_L}\langle 0| O_{\epsilon'}|\epsilon'\rangle_{_R} {}_{_L}\langle \epsilon'| O_{\epsilon'}|0\rangle_{_R} }  } & = 
\frac{  b_{\epsilon'} C_{\alpha \epsilon' \beta} (\frac{2\pi}{L})^{\Delta_{\epsilon'}}+b_{\Box \epsilon'} C_{\alpha ,\Box \epsilon', \beta} (\frac{2\pi}{L})^{\Delta_{\epsilon'}+2}
 +c_{T\overline{T}} C_{\alpha ,T\overline{T}, \beta} (\frac{2\pi}{L})^{\Delta_{T\overline{T}}} + \cdots  }{   b_{\epsilon'} (\frac{2\pi}{L})^{\Delta_{\epsilon'}}+b_{\Box \epsilon'} (\frac{2\pi}{L})^{\Delta_{\epsilon'}+2}
 +c_{T\overline{T}} (\frac{2\pi}{L})^{\Delta_{T\overline{T}}} + \cdots  }
\\
& \approx C_{\alpha \epsilon' \beta} + \frac{e_1}{L^2} + \frac{e_2}{L^{{\Delta_{T\overline{T}}-\Delta_{\epsilon'}}}}+o(\frac{1}{L^4})
\end{aligned}
\end{equation}
Moreover, the extraction of OPE coefficients $C_{\alpha  \epsilon'\alpha}$ is more tricky. Since the operator $\epsilon'$ respects the same symmetry as the original Hamiltonian, all lattice construction of this operators might have non-zero expansion over the identity operators $\hat{\mathbb{I}}$ and $(\hat{T}+\hat{\overline{T}})$. Thus one must linearly combine at least two different lattice operators respecting the same symmetry and construct some combined operator with vanishing expansion over $(\hat{T}+\hat{\overline{T}})$ \cite{Zou2020b,Zou2020a}. Following this idea, we construct other two operators as
\begin{equation}
\begin{aligned}
O_{\epsilon'2} & = \sum_{k_1=1}^{Q-1} \sum_{k_2=1}^{Q-1} [\tau_i^{k_1}  (\sigma_{i+1}^{\dagger} \sigma_{i+2})^{k_2}
 + (\sigma_i^{\dagger} \sigma_{i+1})^{k_1} \tau_{i+2}^{k_2} ]
\\
O_{\epsilon'3} & = \sum_{k_1=1}^{Q-1} \sum_{k_2=1}^{Q-1} [\tau_i^{k_1}  (\sigma_{i+2}^{\dagger} \sigma_{i+3})^{k_2}
 + (\sigma_i^{\dagger} \sigma_{i+1})^{k_1} \tau_{i+3}^{k_2} ]
\end{aligned}
\end{equation}
These two local operators have similar expansion over CFT operators as $O_{\epsilon'1}$ with different coefficients. Then we find the best approximation of CFT operators $\epsilon'$ reads
\begin{equation}
\begin{aligned}
O_A & \approx (0.43+0.10i) \cdot O_{\epsilon'1}+ O_{\epsilon'2}
\\
O_B& \approx O_{\epsilon'2}+ (-0.26+0.07i) \cdot O_{\epsilon'3}
\end{aligned}
\end{equation}
Then the OPE coefficients $C_{\alpha\epsilon'\alpha}$ can be evaluated from
\begin{equation}
\begin{aligned}
&\sqrt{ \frac{  (_{_L}\langle \alpha| O_{A(B)}|\alpha\rangle_{_R} - _{_L}\langle 0| O_{A(B)}|0\rangle_{_R}) (_{_L}\langle \alpha| O_{A(B)}|\alpha\rangle_{_R} - _{_L} \langle 0| O_{A(B)}| 0 \rangle_{_R} ) }{ _{_L}\langle 0| O_{A(B)}|\epsilon'\rangle_{_R} {}_{_L}\langle \epsilon'| O_{A(B)}|0\rangle_{_R} }  } 
\\
 = & 
\frac{  b_{\epsilon'} C_{\alpha \epsilon' \alpha} (\frac{2\pi}{L})^{\Delta_{\epsilon'}}+b_{\Box \epsilon'} C_{\alpha ,\Box \epsilon', \alpha} (\frac{2\pi}{L})^{\Delta_{\epsilon'}+2}
 +c_{T\overline{T}} C_{\alpha ,T\overline{T}, \alpha} (\frac{2\pi}{L})^{\Delta_{T\overline{T}}} + \cdots  }{   b_{\epsilon'} (\frac{2\pi}{L})^{\Delta_{\epsilon'}}+b_{\Box \epsilon'} (\frac{2\pi}{L})^{\Delta_{\epsilon'}+2}
 +c_{T\overline{T}} (\frac{2\pi}{L})^{\Delta_{T\overline{T}}} + \cdots  }
\\
 \approx & C_{\alpha \epsilon' \alpha} + \frac{e_1}{L^2} + \frac{e_2}{L^{{\Delta_{T\overline{T}}-\Delta_{\epsilon'}}}}+o(\frac{1}{L^4})
\end{aligned}
\end{equation}
We have subtracted each local correlators with its expectation value on the ground states to remove the contribution from identity operator. 

The numerical results of different OPE coefficients are shown in Fig~\ref{fig:scale_ope}.
Generally, the finite-size scaling errors of the imaginary parts are larger than those of the real parts. 
We note that most of the scaling functions are non-monotonic, which come from the complex values of scaling dimensions $\Delta_n$, an intrinsic property of complex CFT. The fitting process using non-monotonic functions may be one of possible reason to observe large fitting errors.

To get a reliable estimation of OPE coefficients, our strategy is, for each OPE coefficient we select at least two local spin operators to compute the correlation functions. The results from different local operators cross-check the correctness of the obtained result, and provide a way to estimate the extrapolated errors. In this regard, we use the mean value of different local operators as the estimation of OPE coefficients and the associated error comes from the difference between the mean value and the results from different local operators.   
The final results shown in Tab. II in the main text.

\begin{figure*}
\includegraphics[width=0.27\textwidth]{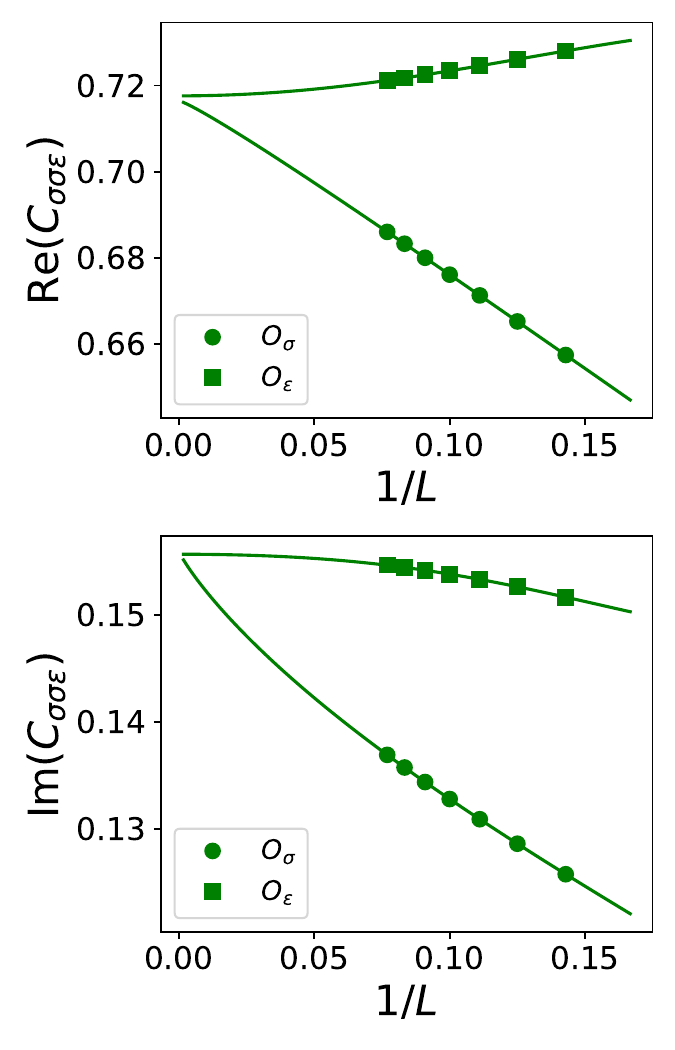}
\includegraphics[width=0.27\textwidth]{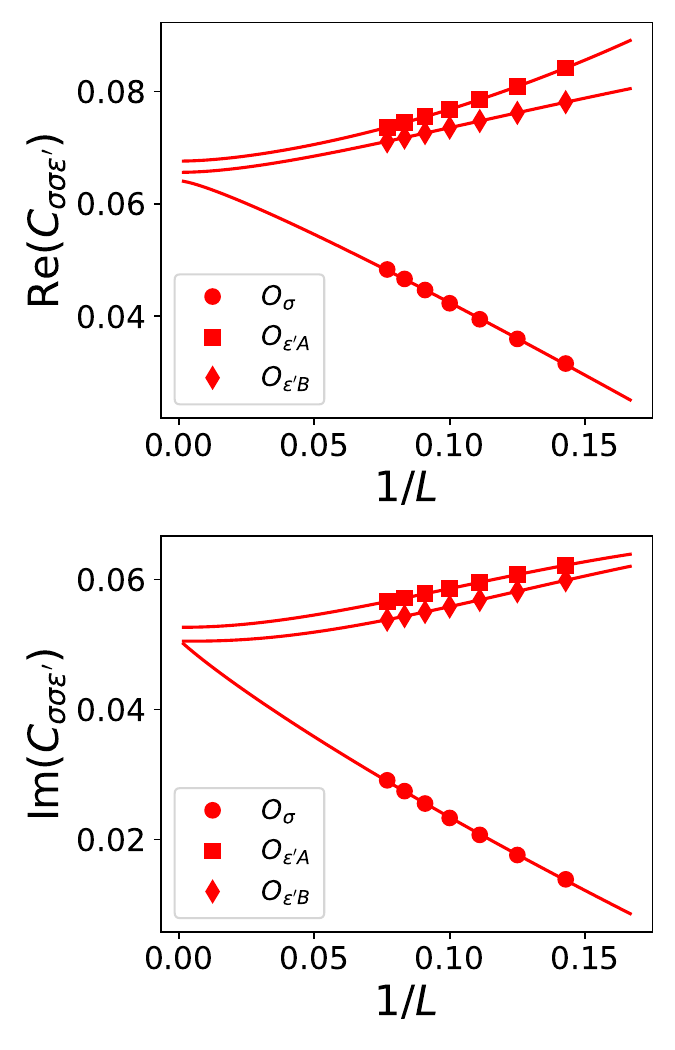}
\includegraphics[width=0.27\textwidth]{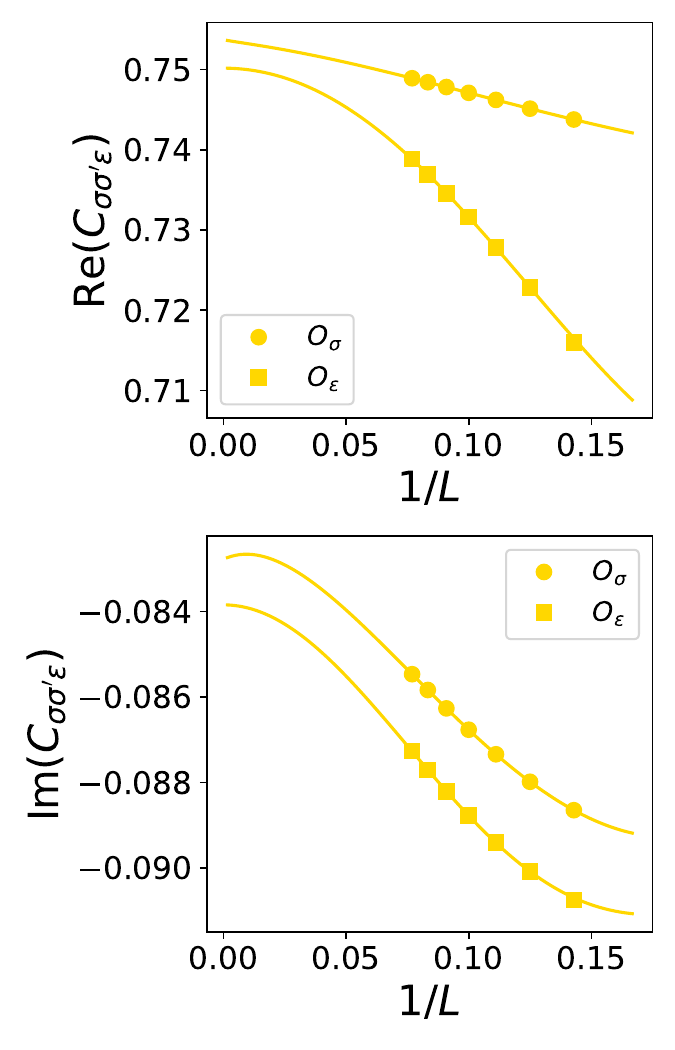}
\includegraphics[width=0.27\textwidth]{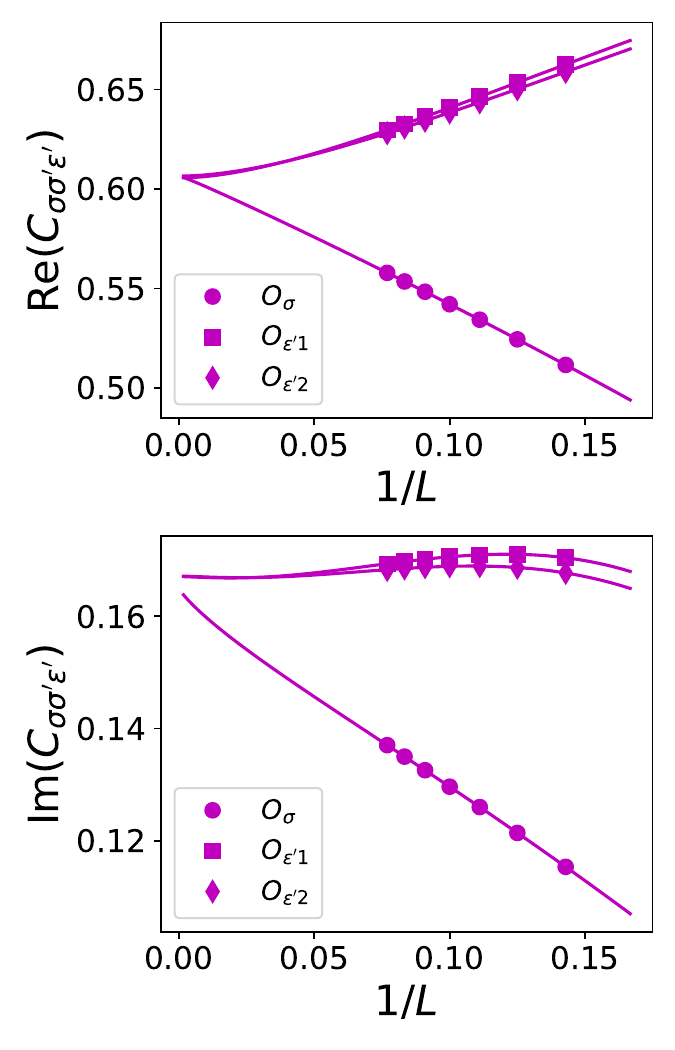}
\includegraphics[width=0.27\textwidth]{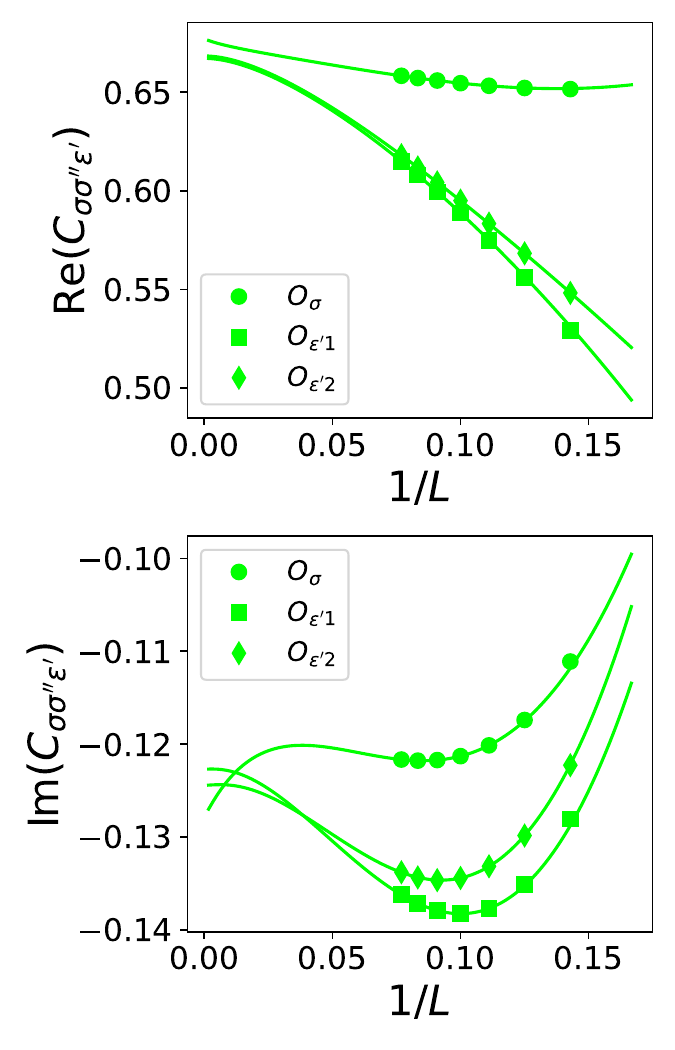}
\includegraphics[width=0.27\textwidth]{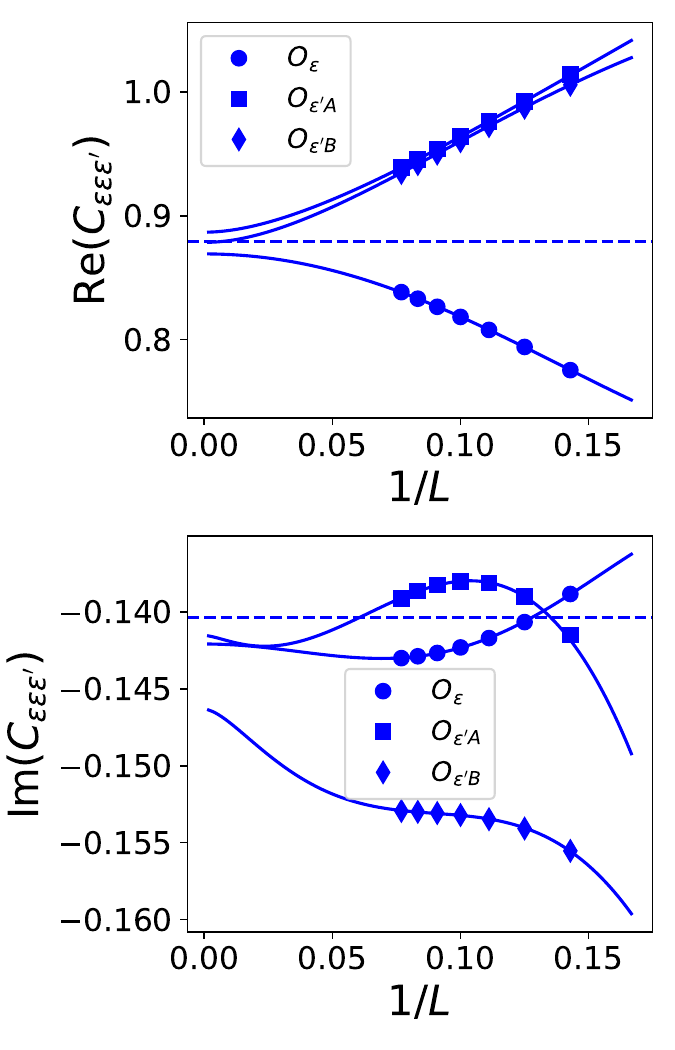}
\includegraphics[width=0.27\textwidth]{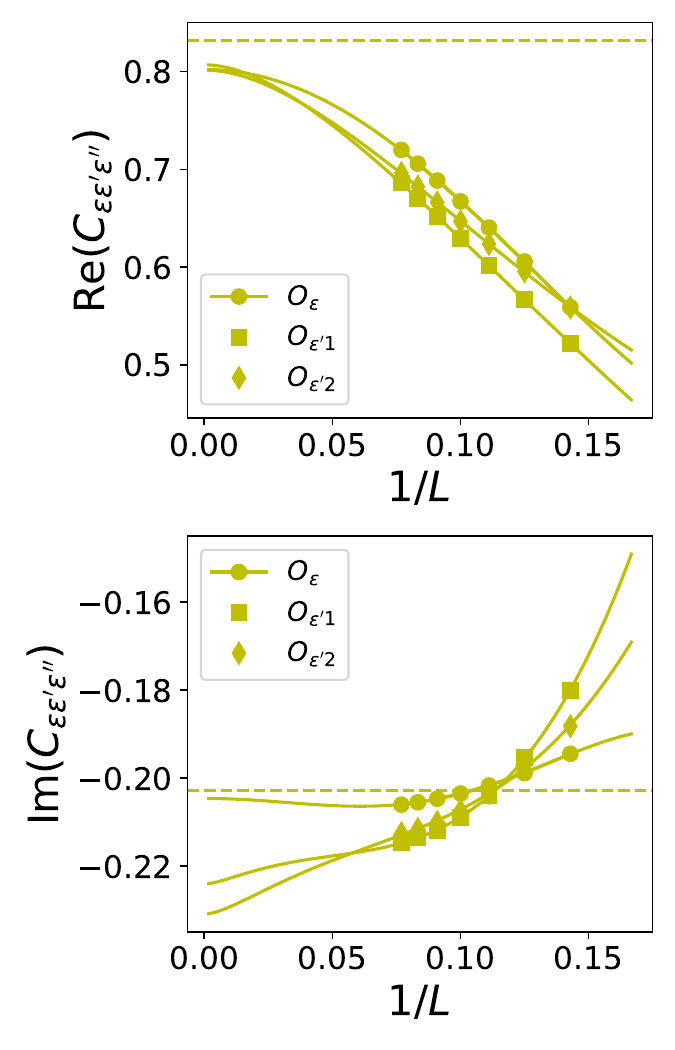}
\includegraphics[width=0.27\textwidth]{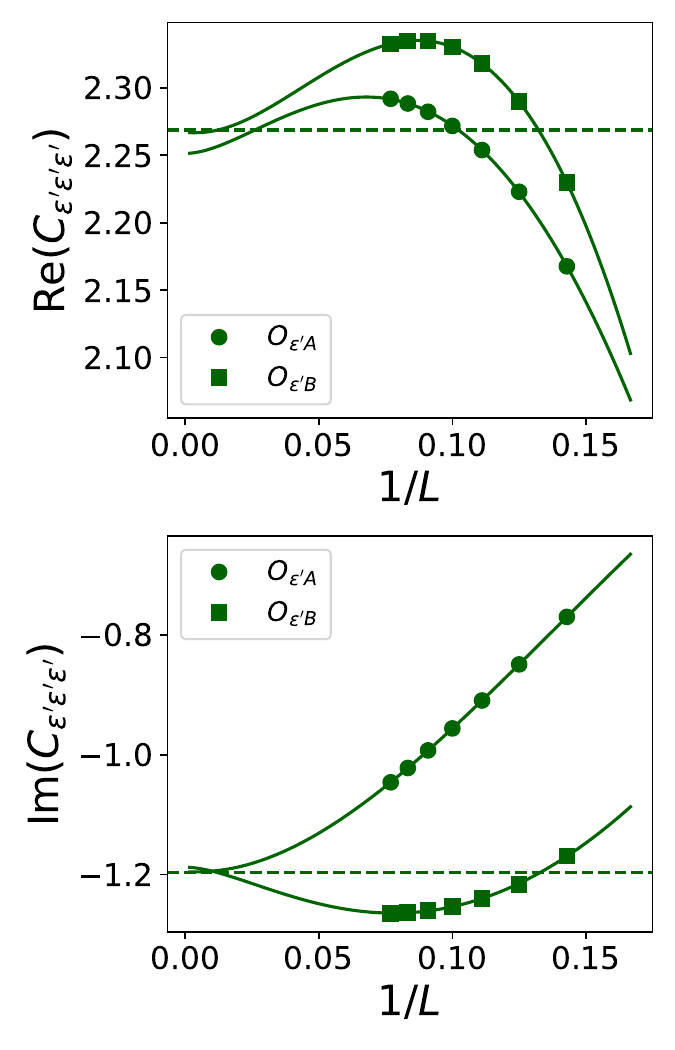}
\includegraphics[width=0.27\textwidth]{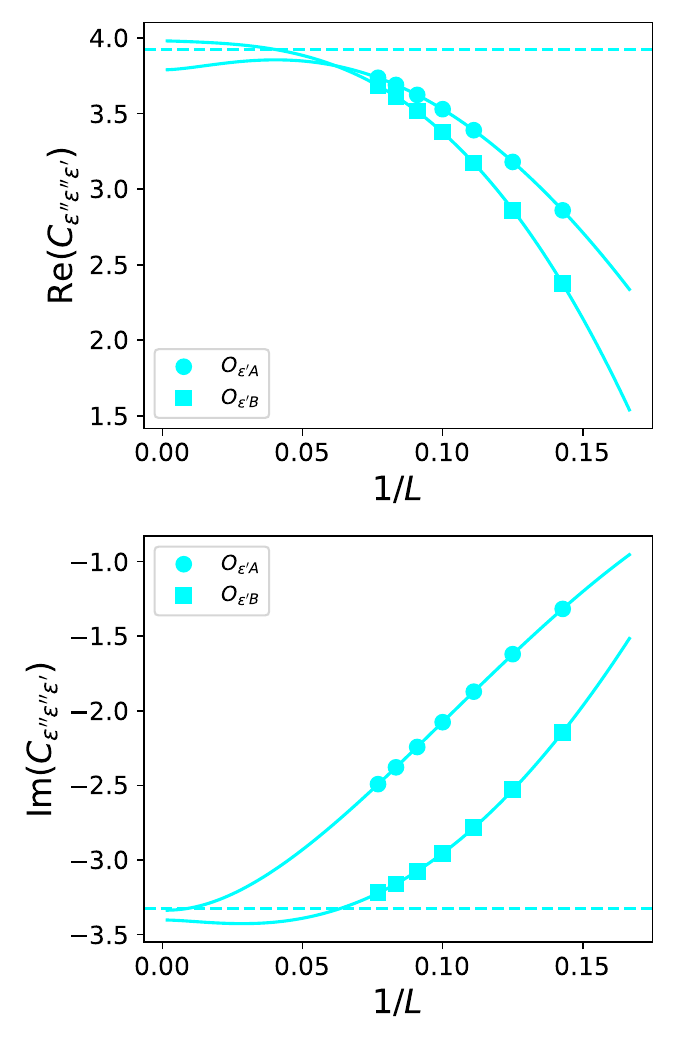}
\caption{\label{fig:scale_ope} The OPE coefficients obtained from the finite-size extrapolation of the specific scaling functions (see Eq. S35, S36, S37, S40). The real and imaginary part of the OPE coefficients are present in separated subfigures. The different symbols represent the result from different local lattice operators. Some evaluation of OPE coefficients from analytical continuation \cite{Poghossian2013} are shown by dashed lines.}
\end{figure*}

\begin{table}[!htb]
\caption{\label{tab:con_familiy_epsilon} Conformal multiplet of $\epsilon$. The numerical results calculated from our model $H_{\text{NH-Potts}}(J_c,h_c,\lambda_c)$ are compared with analytical prediction from Coulomb Gas partition function \cite{Gorbenko2018b}. $L_{-n}/\overline{L}_{-n}$ represent Virasoro generators connecting different states (the same applies to following tables).}

\begin{tabular}{c|c|c|c}

\hline
\hline
 Operators & s &  complex CFT & non-Hermitian 5-Potts  \\ \hline
 
$\epsilon$ & 0 &  $0.4656-0.2245i$ &  $0.4633-0.2240i$ \\ \hline

$L_{-1} \epsilon$ & 1 &  $1.4656-0.2245i$ & $1.4623-0.2218i$ \\ \hline

$L^2_{-1} \epsilon$ & 2 &   $2.4656-0.2245i$ & $2.4629-0.2099i$ \\  \hline

$L_{-1} \overline{L}_{-1} \epsilon$ & 0&  $2.4656-0.2245i$ & $2.4696-0.2158i$ \\  \hline

$L^3_{-1} \epsilon$ & 3 &   $3.4656-0.2245i$ & $3.4611-0.2279i$ \\ 
$L_{-3} \epsilon$ & 3 &  $3.4656-0.2245i$ & $3.4612-0.1695i$ \\  \hline

$L^2_{-1} \overline{L}_{-1} \epsilon$ & 1 &   $3.4656-0.2245i$ & $3.4821-0.1985i$\\  \hline

$L^3_{-1} \overline{L}_{-1}  \epsilon$ & 2 &   $4.4656-0.2245i$ & $4.4753-0.2174i$ \\ 
$L_{-3} \overline{L}_{-1}  \epsilon$ & 2 & $4.4656-0.2245i$  & $4.5017-0.1411i$ \\  \hline

$L^2_{-1} \overline{L}^2_{-1} \epsilon$ & 0 &   $4.4656-0.2245i$ & $4.5147-0.1611i$\\ \hline
\hline
\end{tabular}
\end{table}

\begin{table}
\caption{\label{tab:con_familiy_epsilon'} Conformal multiplet of $\epsilon '$.}

\begin{tabular}{c|c|c|c}
\hline
\hline
 Operators & s & complex CFT & non-Hermitian 5-Potts  \\ \hline
 
$\epsilon'$ & 0 & $1.9083-0.5987i$ &  $1.8998-0.5977i$ \\ \hline

$L_{-1} \epsilon'$ & 1 &  $2.9083-0.5987i$ & $2.9097-0.5901i$ \\ \hline

$L^2_{-1} \epsilon'$ & 2 &  $3.9083-0.5987i$ & $3.9099-0.5874i$\\
$L_{-2} \epsilon'$ & 2 & $3.9083-0.5987i$ & $3.9121-0.5626i$ \\ \hline

$L_{-1} \overline{L}_{-1} \epsilon'$ & 0&  $3.9083-0.5987i$ & $3.9219-0.5756i$ \\ \hline

$L^2_{-1} \overline{L}_{-1} \epsilon'$ & 1 &   $4.9083-0.5987i$ & $4.9376-0.5664i$\\ 
$L_{-2} \overline{L}_{-1} \epsilon'$ & 1 & $4.9083-0.5987i$ & $4.9448-0.5317i$\\ \hline

$L^3_{-1}  \epsilon'$ & 3 &   $4.9083-0.5987i$ & $4.9117-0.4825i$ \\
$L_{-1} L_{-2} \epsilon'$& 3 & $4.9083-0.5987i$ & $4.9199-0.5631i$ \\ \hline
\hline
\end{tabular}
\end{table}

\begin{table}
\caption{\label{tab:con_familiy_sigma} Conformal multiplet of $\sigma$.}
\begin{tabular}{c|c|c|c}
\hline
\hline
 Operators & s & complex CFT & non-Hermitian 5-Potts  \\ \hline
 
$\sigma$ & 0 & $0.1336-0.0205i$ &  $0.1334-0.0206i$ \\ \hline

$L_{-1} \sigma$ & 1 &  $1.1336-0.0205i$ &  $1.1328-0.0206i$\\ \hline

$L^2_{-1}  \sigma$ & 2 &  $2.1336-0.0205i$ & $2.1317-0.0174i$\\
$L_{-2}  \sigma$ &2 & $2.1336-0.0205i$ & $2.1329-0.0224i$ \\ \hline

$L_{-1} \overline{L}_{-1} \sigma$ & 0&  $2.1336-0.0205i$ & $2.1363-0.0194i$ \\ \hline

$L^3_{-1}  \sigma$ & 3 &  $3.1336-0.0205i$ &  $3.1231+0.0024i$\\
$L_{-1} L_{-2}  \sigma$ & 3 & $3.1336-0.0205i$ & $3.1280-0.0181i$\\
$L_{-3} \sigma$ & 3 & $3.1336-0.0205i$ & $3.1321-0.0198i$\\ \hline

$L^2_{-1} \overline{L}_{-1} \sigma$ & 1 &  $3.1336-0.0205i$ & $3.1356-0.0199i$\\
$L_{-2} \overline{L}_{-1} \sigma$ & 1 & $3.1336-0.0205i$ & $3.1439-0.0123i$ \\ \hline

$L^3_{-1} \overline{L}_{-1} \sigma$ & 2 &  $4.1336-0.0205i$ & $4.1391-0.0035i$\\
$L_{-1} L_{-2} \overline{L}_{-1}  \sigma$ & 2 & $4.1336-0.0205i$ & $4.1459-0.0158i$\\
$L_{-3} \overline{L}_{-1} \sigma$ & 2 & $4.1336-0.0205i$ & $4.1534+0.0208i$\\ \hline

$L^2_{-1} \overline{L}^2_{-1} \sigma$ & 0 &  $4.1336-0.0205i$ &  $4.1452-0.0061i$\\
$L_{-2} \overline{L}^2_{-1} \sigma$ & 0 & $4.1336-0.0205i$ & $4.1482-0.0034i$\\
$L^2_{-1} \overline{L}_{-2} \sigma$ & 0 & $4.1336-0.0205i$ & $4.1502-0.0205i$\\
$L_{-2} \overline{L}_{-2} \sigma$ & 0 & $4.1336-0.0205i$ & $4.1643+0.0065i$\\ \hline
\hline
\end{tabular}
\end{table}

\begin{table}
\caption{\label{tab:con_familiy_sigma'} Conformal multiplet of $\sigma'$.}
\begin{tabular}{c|c|c|c}
\hline
\hline
 Operators & s & complex CFT & non-Hermitian 5-Potts  \\ \hline
 
$\sigma'$ & 0 & $1.1107-0.1701i$ &  $1.1081-0.1702i$ \\ \hline

$L_{-1} \sigma'$ & 1 &  $2.1107-0.1701i$ & $2.1093-0.1675i$\\ \hline

$L^2_{-1}  \sigma'$ & 2 &  $3.1107-0.1701i$ & $3.1117-0.1649i$ \\
$L_{-2} \sigma'$ & 2 & $3.1107-0.1701i$ & $3.1122-0.1521i$\\ \hline

$L_{-1} \overline{L}_{-1} \sigma'$ & 0&  $3.1107-0.1701i$ & $3.1186-0.1604i$\\ \hline

$L^3_{-1} \sigma'$ & 3 &  $4.1107-0.1701i$ & $4.1057-0.1023i$ \\
$L_{-1} L_{-2} \sigma'$ & 3 & $4.1107-0.1701i$ & $4.1065-0.1746i$\\
$L_{-3} \sigma'$ & 3 & $4.1107-0.1701i$ & $4.1083-0.1480i$\\ \hline

$L^2_{-1} \overline{L}_{-1} \sigma'$ & 1 & $4.1107-0.1701i$ & $4.1268-0.1550i$\\
$L_{-2} \overline{L}_{-1} \sigma'$ & 1 & $4.1107-0.1701i$ & $4.1340-0.1335i$\\ \hline
\hline
\end{tabular}
\end{table}

\begin{table}
\caption{\label{tab:con_familiy_sigma''} Conformal multiplet of $\sigma ''$.}
\begin{tabular}{c|c|c|c}
\hline
\hline
 Operators & s & complex CFT & non-Hermitian 5-Potts  \\ \hline
 
$\sigma''$ & 0 & $3.0648-0.4695i$ &  $3.0647-0.4702i$ \\ \hline

$L_{-1} \sigma''$ & 1 & $4.0648-0.4695i$ & $4.0720-0.4550i$\\ \hline
\hline
\end{tabular}
\end{table}

\begin{table}
\caption{\label{tab:con_familiy_Z} Conformal multiplet of $Z$.}
\begin{tabular}{c|c|c|c}
\hline
\hline
 Operators & s & complex CFT & non-Hermitian 5-Potts  \\ \hline
 
$Z$ & 0 & $2.0115+0.3046i$ &  $2.0162+0.3042i$ \\ \hline

$L_{-1} Z$ & 1 &  $3.0115+0.3046i$ & $3.0215+0.3085i$\\ \hline

$L^2_{-1} Z$ & 2 &  $4.0115+0.3046i$ &$4.0210+0.3292i$ \\
$L_{-2} Z$ & 2 & $4.0115+0.3046i$ &  $4.0251+0.3070i$\\ \hline

$L_{-1} \overline{L}_{-1} Z$ & 0&  $4.0115+0.3046i$ & $4.0332+0.3193i$\\ \hline
\hline
\end{tabular}
\end{table}

\begin{table}
\caption{\label{tab:con_familiy_X} Conformal multiplet of $X$.}
\begin{tabular}{c|c|c|c}
\hline
\hline

 Operators & s & complex CFT & non-Hermitian 5-Potts  \\ \hline
 
$X$ & 1 & $2.1336+0.2859i$ & $2.1378+0.2858i$\\ \hline

$L_{-1} X$ & 2 &  $3.1336+0.2859i$ & $3.1397+0.2912i$\\ \hline

$\overline{L}_{-1} X$ & 0 &  $3.1336+0.2859i$ & $3.1449+0.2886i$\\ 
$L_{-1} \overline{X}$ & 0 &  $3.1336+0.2859i$ & $3.1449+0.2886i$\\ 
\hline

$L^2_{-1}  X$ & 3&  $4.1336+0.2859i$ & $4.1273+0.3131i$ \\
$L_{-2}  X$ & 3 & $4.1336+0.2859i$ & $4.1426+0.2910i$\\ \hline

$L_{-1} \overline{L}_{-1} X$ & 1&  $4.1336+0.2859i$ & $4.1492+0.2856i$ \\
$L^2_{-1} \overline{X}$ & 1 & $4.1336+0.2859i$ & $4.1529+0.3074i$\\
$L_{-2} \overline{X} $ & 1 & $4.1336+0.2859i$ & $4.1538+0.3015i$\\ \hline
\hline
\end{tabular}
\end{table}

\begin{table}
\caption{\label{tab:con_familiy_Y} Conformal multiplet of $Y$.}
\begin{tabular}{c|c|c|c}
\hline
\hline
 Operators & s & complex CFT & non-Hermitian 5-Potts  \\ \hline
 
$Y$ & 2 & $2.5+0.2298i$ &  $2.5033+0.2300i$ \\ \hline

$L_{-1} Y$ & 3 &  $3.5+0.2298i$ & $3.4982+0.2363i$ \\ \hline

$\overline{L}_{-1} Y$ & 1 &  $3.5+0.2298i$ & $3.5110+0.2315i$\\ \hline

$\overline{L}^2_{-1}  Y$ & 0&  $4.5+0.2298i$ & $4.4407+0.2405i$\\
$\overline{L}_{-2}  Y$ & 0 & $4.5+0.2298i$ & $4.5216+0.2481i$\\
$L^2_{-1}  \overline{Y}$ & 0 & $4.5+0.2298i$ & $4.5255+0.2478i$\\
$L_{-2} \overline{Y} $ & 0 & $4.5+0.2298i$ & $4.6079+0.2449i$\\ \hline

$L_{-1} \overline{L}_{-1} Y$ & 2&  $4.5+0.2298i$ & $4.5151+0.2480i$ \\ \hline
\hline

\end{tabular}
\end{table}

\begin{table}
\caption{\label{tab:con_familiy_W} Conformal multiplet of $W$.}
\begin{tabular}{c|c|c|c}
\hline
\hline
 Operators & s & complex CFT & non-Hermitian 5-Potts  \\ \hline
 
$W$ & 3 & $3.1107+0.1362i$ &  $3.1102+0.1374i$ \\ \hline

$\overline{L}_{-1} W$ & 2 &  $4.1107+0.1362i$ & $4.1197+0.1373i$\\ \hline
\hline
\end{tabular}
\end{table}

\begin{table}
\caption{\label{tab:con_familiy_T} Conformal multiplet of $T$.}
\begin{tabular}{c|c|c|c}
\hline
\hline
 Operators & s & complex CFT & non-Hermitian 5-Potts  \\ \hline
 
$L_{-1} T$ & 3 &  $3.0+0.0i$ & $2.9983+0.0098i$ \\ \hline

$T\overline{T}$ & 0 & $4.0+0.0i$ &  $4.0159+0.0057i$ \\ \hline
\hline
\end{tabular}
\end{table}

\end{widetext}

\end{document}